\documentclass{pasj00}
\draft

\begin{document}
\SetRunningHead{Kaneda et al.}{PAHs in molecular loops}
\Received{//}
\Accepted{//}

\title{Processing of polycyclic aromatic hydrocarbons in molecular-loop regions near the Galactic center revealed by AKARI}

\author{%
   Hidehiro \textsc{Kaneda},\altaffilmark{1}
   Daisuke \textsc{Ishihara},\altaffilmark{1}
   Akio \textsc{Mouri},\altaffilmark{1}\\  
   Shinki \textsc{Oyabu},\altaffilmark{1}
   Mitsuyoshi \textsc{Yamagishi},\altaffilmark{1}
   Toru \textsc{Kondo},\altaffilmark{1}
   Takashi \textsc{Onaka},\altaffilmark{2}\\
   Yasuo \textsc{Fukui},\altaffilmark{1}
   Akiko \textsc{Kawamura},\altaffilmark{1}
   Kazufumi \textsc{Torii}\altaffilmark{1}
}
 \altaffiltext{1}{Graduate School of Science, Nagoya University \\
Nagoya, Aichi 464-8602}
 \email{kaneda@u.phys.nagoya-u.ac.jp}
\altaffiltext{2}{Department of Astronomy, Graduate School of Science, University of Tokyo, \\
Bunkyo-ku, Tokyo 113-0033}



\maketitle

\begin{abstract}

We present the AKARI mid-infrared diffuse map of an area of about $4^{\circ}\times3^{\circ}$ near the Galactic center in the 9 $\mu$m band. The band intensity is mostly attributed to the aromatic hydrocarbon infrared emissions of carbonaceous grains at wavelengths of 6.2, 7,7, 8.6, and 11.3 $\mu$m. We detect the 9 $\mu$m emission structures extending from the Galactic plane up to the latitude of $\sim\timeform{2^{\circ}.5}$, which have spatial correspondence to the molecular loops revealed by the NANTEN $^{12}$CO $(J=1-0)$ observations. We find that the surface brightness at 9 $\mu$m is suppressed near the foot points of the CO loops. The ratios of the 9 $\mu$m to the IRAS 100 $\mu$m brightness show significant depression near such bright regions in the CO emission. With the AKARI near-IR 2.5--5 $\mu$m spectroscopy, we find that the 3.3 $\mu$m aromatic hydrocarbon emission is absent in the region associated with the loop. These suggest the processing and destruction of carbonaceous grains in the CO molecular loops. 

\end{abstract}

\section{Introduction}

Many past and current infrared (IR) observatories, both space-based and ground-based, have revealed that very small carbonaceous grains including polycyclic aromatic hydrocarbons (PAHs) are ubiquitous in the general interstellar media (ISM) in our Galaxy and nearby galaxies (e.g. Mattila et al. 1996; Onaka et al. 1996; Abergel et al. 2002; Smith et al. 2007). The types of the galaxies span a wide range even including elliptical galaxies (Kaneda et al. 2005, 2008) and radio galaxies \citep{Stu06, Lei09}. Among various gas phases of the ISM, PAHs are associated most tightly with neutral atomic and molecular gas, while their emission is relatively weak in ionized regions with strong radiation field probably due to destruction through the absorption of ultraviolet (UV) light \citep{Bou88,Des90,Tie08}. It is observationally known that the PAH emission is well correlated with far-IR cold dust emission but rather poorly correlated with warm dust emission \citep{Ben08}. The cold and warm dust components are related to the total gas content and star-formation activity, respectively \citep{Cox89, Sau05, Suz10}. Thus PAHs are excellent tracers of the total amount of the general ISM. Their advantage over the cold dust is spatial resolution much better in the near- to mid-IR PAH emission than in the far-IR dust emission, because the imaging performance of an infrared telescope is generally limited by diffraction, which is severer as the observed wavelength is longer.  

Despite the ubiquity as a whole, PAHs are very fragile and expected to be processed and destroyed quite easily through collisions in local interstellar shock regions and hot plasma. The collisional destruction can be far more effective than the UV photo-destruction (e.g. Tielens 2008). With the advent of Spitzer and AKARI, which possess unprecedentedly high sensitivity for the PAH emission, it is found that the PAH emission is indeed highly suppressed as compared to dust continuum emission in supernova remnants with such hostile conditions (e.g. Reach et al. 2006; Ishihara et al. 2010b).
From recent theoretical studies combined with laboratory data, Micelotta et al. (2010a, 2010b) calculated the collisional destruction efficiency of PAHs in interstellar shocks and hot plasma. According to their results, for shock velocities higher than $\sim$100 km s$^{-1}$, PAHs are completely destroyed in postshock hot plasma. It should be noted that the lifetime of dust grains is $10^2-10^3$ times longer than that of PAHs even if grains are as small as PAHs. This is because the dissociation yields of 2-dimentional PAHs are close to unity, while the sputtering yields of 3-dimensional dust grains are much smaller than unity \citep{Mic10a}. Since the sizes of dust responsible for far-IR continuum emission are $10^2-10^3$ times as large as those of typical PAHs ($\sim 6$ \AA; Tielens 2008), far-IR dust can survive against the shock on time scales $10^4-10^6$ times longer than PAHs.

On the other hand, for shock velocities significantly lower than 100 km s$^{-1}$, PAHs in shocks are not completely destroyed but their structures are significantly affected. In such a slow shock, PAHs, or small hydrogenated amorphous carbons, may even be formed by shattering of larger carbonaceous grains \citep{Jon96}. Such processing of PAHs, or structural changes of hydrocarbons, can be traced by near-IR spectroscopy for the 3.3 $\mu$m main feature and 3.4--3.6 $\mu$m sub-features, both of which are attributed to the C-H vibration mode of carbonaceous grains. The former is due to aromatic ($sp^2$) hydrocarbons, while the latter is likely attributed to aliphatic ($sp^3$) hydrocarbons \citep{Dul81}. The sub-features can also be caused by emission from upper vibrational levels of excited PAHs because of anharmonicity \citep{Bar87}. The aromatic feature at 3.3 $\mu$m is sensitive to smallest PAHs \citep{Sch93}, and hence its intensity relative to those at longer wavelengths provides information on the size distribution of PAHs. Laboratory experiments showed that the relative strength of the 3.4 $\mu$m over the 3.3 $\mu$m band decreased after thermal annealing \citep{Sak90,Got00}, indicating the structural change of carbonaceous grains from aliphatic to aromatic. Thus the relative strength of the 3.3 $\mu$m main feature to the sub-feature also provides information on the history of the evolution of carbonaceous grains in the interstellar environment. 

By utilizing the nature that PAHs are very easily processed and destroyed in harsh interstellar environments, we investigate the condition of molecular clouds near the Galactic center. In particular, we focus on the two molecular loops, loops 1 and 2, found by the $^{12}$CO $(J=1-0)$ emission with the NANTEN telescope \citep{Fuk06}. The molecular loops are located about 300 pc away from the Galactic center. \citet{Fuk06} suggested that such molecular loops were formed by the buoyant rise of magnetic loops due to magnetic flotation driven by the Parker instability (Parker 1966). The numerical simulation predicted that the floated gas slides down along the loop and then collides with disk gas, forming dense structures at the foot points of the loops \citep{Mat88}. If the fall velocity exceeds the sound velocity, shock must be generated in the cloud. This phenomenon has a possibility to give some clue to long-standing questions such as an origin of high kinetic temperatures of molecular clouds in the Galactic center. The several follow-up studies performed on the basis of CO observations (Torii et al. 2010a, 2010b; Fujishita et al. 2009; Kudo et al. 2011) and numerical simulations \citep{Mac09,Tak09} consistently supported the above picture. 

We present the AKARI mid-IR 9 $\mu$m map near the Galactic center region to show the global distribution of the PAH emission. We find extraplanar PAH emission structures extended from the Galactic plane, which have spatial correspondence to the above molecular loops. We also present the result of the $2.5-5$ $\mu$m spectroscopic observations of the foot point and the top of loop 2 to show the properties of the PAHs therein. 
Our observational results suggest the processing and destruction of carbonaceous grains in the molecular loops, which may provide another piece of observational evidence for the above phenomena from an entirely different aspect from the CO observations. 

\section{Data reduction}
AKARI carried out all-sky surveys in the six photometric bands in the mid- and far-IR during the cold mission phase (Phases 1 and 2) with the cryogenic telescope cooled to 6 K by liquid helium \citep{Mur07,Ona07,Kaw07}. In this paper, among them, we use a part of the 9 $\mu$m broad-band survey map, which is of the far highest quality as of writing the paper \citep{Ish10a}. We removed cosmic-ray-induced spikes and changes in the offset of detector output with the algorithms developed by Ishihara et al. (2006, 2010a) and \citet{Mou11}, respectively. We combined data taken in different seasons by removing the seasonal variation of the Zodiacal light emission \citep{Mou11}. Details of the map-making method will be described in another paper (Ishihara et al. in prep.). The spatial bin size of the original 9 $\mu$m map data is $\timeform{9.36''}$, limited by the telemetry downlink rate, while the full-width at half-maximum of the point spread function measured in pointed imaging observations is $\sim\timeform{5.5''}$ at 9 $\mu$m \citep{Ona96}. We regridded the map to the bin size of $12''$. After the detection and removal of point-like sources, we smoothed the map to the spatial resolution of $2'$ in order to ensure signal-to-noise ratios higher than 5 for diffuse emission at high Galactic latitudes; the resolution size is still 1--2 orders of magnitude smaller than the spatial scales of the molecular loop structures to be discussed below. 

The resultant map of an area of about $4^{\circ}\times3^{\circ}$ near the Galactic center is shown in figure 1. The constant sky background level estimated from regions at the Galactic latitudes of $\sim\pm 3^{\circ}$ is subtracted. Although the data have been corrected for non-linearity to $<2$ \% levels for the range of the surface brightness treated in the present study \citep{Kat10, Ish10a}, a linear conversion factor of an analog-to-digital unit (ADU) to MJy sr$^{-1}$ is yet to be finalized for diffuse maps, as of writing the paper. Below, if necessary, we apply a preliminary conversion factor to convert ADU to MJy sr$^{-1}$ for the 9 $\mu$m map (Ishihara et al. in prep.). However most of our analyses and results do not require information on the absolute surface brightness in the physical units. Therefore, unless explicitly stated, we use ADUs or normalized units for the 9 $\mu$m surface brightness.

In the AKARI warm mission phase (Phase 3) after the boil-off of the liquid-helium cryogen, we performed near-IR slit-spectroscopic observations targeted at various positions in our Galaxy. In this paper, we present four spectra obtained from on-plane and off-plane molecular loop regions near the Galactic center. To remove hot pixels due to high operating temperature from the array images, we applied the same custom procedure as described in \citet{Yam11} in addition to the basic pipeline process, which resulted in degrading the resolution of the spectra by 3 spectral bins. In many cases, Galactic discrete sources are found to be present in the slit apertures, which make it difficult to extract the spectra of pure Galactic diffuse emission. Below we used only the slit data which included 10 continuous spatial bins free of emission from discrete sources and summed the slit-aperture data of such 10 spatial bins to create diffuse spectra. For comparative purpose, in figure 2, we show the spectrum of typical Galactic diffuse emission taken in the AKARI cold mission phase; this was selected because the data were least contaminated by discrete sources among those taken from inner Galactic regions.  
The spectrum has significantly better quality, especially higher spectral resolution and lower noise thanks to lower operating temperature than the spectra taken in Phase 3. The spectral features seen at wavelengths of 3.3, 6.2, 7,7, 8.6, and 11.3 $\mu$m originate from the aromatic hydrocarbon infrared bands (e.g., Allamandola et al. 1989). The log of these spectroscopic observations is listed in table 1. The two observations of the IDs starting with `52' were performed during AKARI Director's Time; we observed the same positions three times to improve data quality.  The others were carried out in part of the AKARI mission program ``ISM in our Galaxy and Nearby galaxies'' (ISMGN; Kaneda et al. 2009). `Ns' and `Nh' are different slit apertures with sizes of $5''\times\timeform{0.8'}$ and $3''\times1'$, respectively \citep{Ona07}. 

\begin{table*}
\caption{Observation Log}
\begin{center}
\begin{tabular}{lcccc}
\hline\hline
Position ($l$, $b$)\footnotemark[$*$]&Date&Slit&Observation ID&Notes\\ \hline
$l=\timeform{311.50^{\circ}}$, $b=\timeform{0.50^{\circ}}$&Feb 12, 2007&Ns&1400259&figures 2 and 7e\\
$l=\timeform{356.06^{\circ}}$, $b=\timeform{0.03^{\circ}}$&Sep 18, 2008&Nh&1420697&figures 6a and 7a\\
$l=\timeform{356.61^{\circ}}$, $b=\timeform{0.28^{\circ}}$&Sep 17, 2008&Nh&1420698&figures 6b and 7b\\
$l=\timeform{355.36^{\circ}}$, $b=\timeform{0.92^{\circ}}$&Sep 17, 2009&Ns&5201036&figures 6c and 7c\\
$l=\timeform{355.27^{\circ}}$, $b=\timeform{1.48^{\circ}}$&Sep 16, 2009&Ns&5201037&figures 6d and 7d\\
\hline
\\
\multicolumn{5}{@{}l@{}}{\hbox to 0pt{\parbox{130mm}{\footnotesize

\par\noindent
\footnotemark[$*$] Central positions of the sub-slit apertures. $l$ and $b$ denotes the Galactic longitude and latitude, respectively, in the Galactic coordinates.  }}}
\end{tabular}
\end{center}
\end{table*}

\begin{figure}
\FigureFile(85mm,85mm){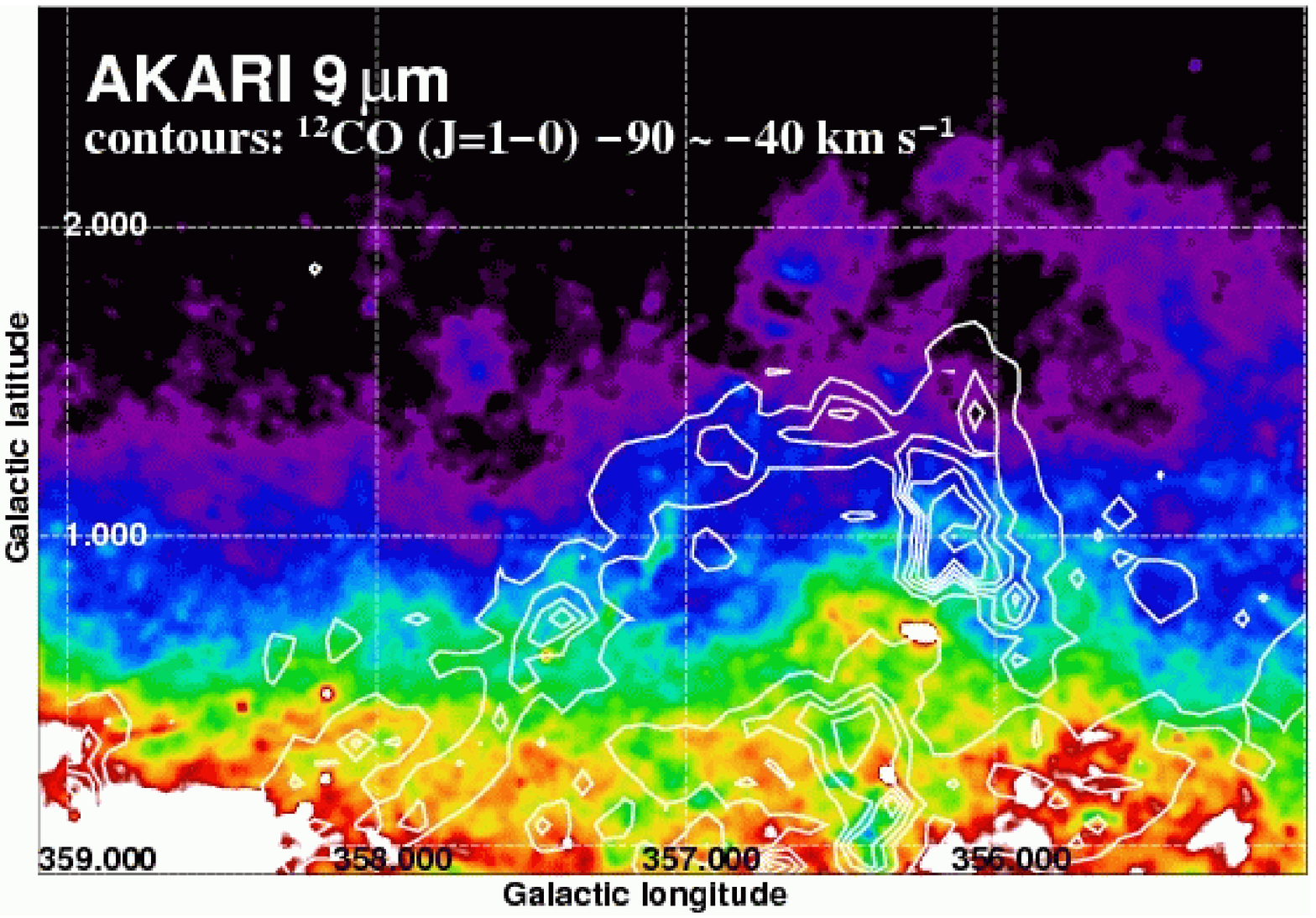}\quad\FigureFile(85mm,85mm){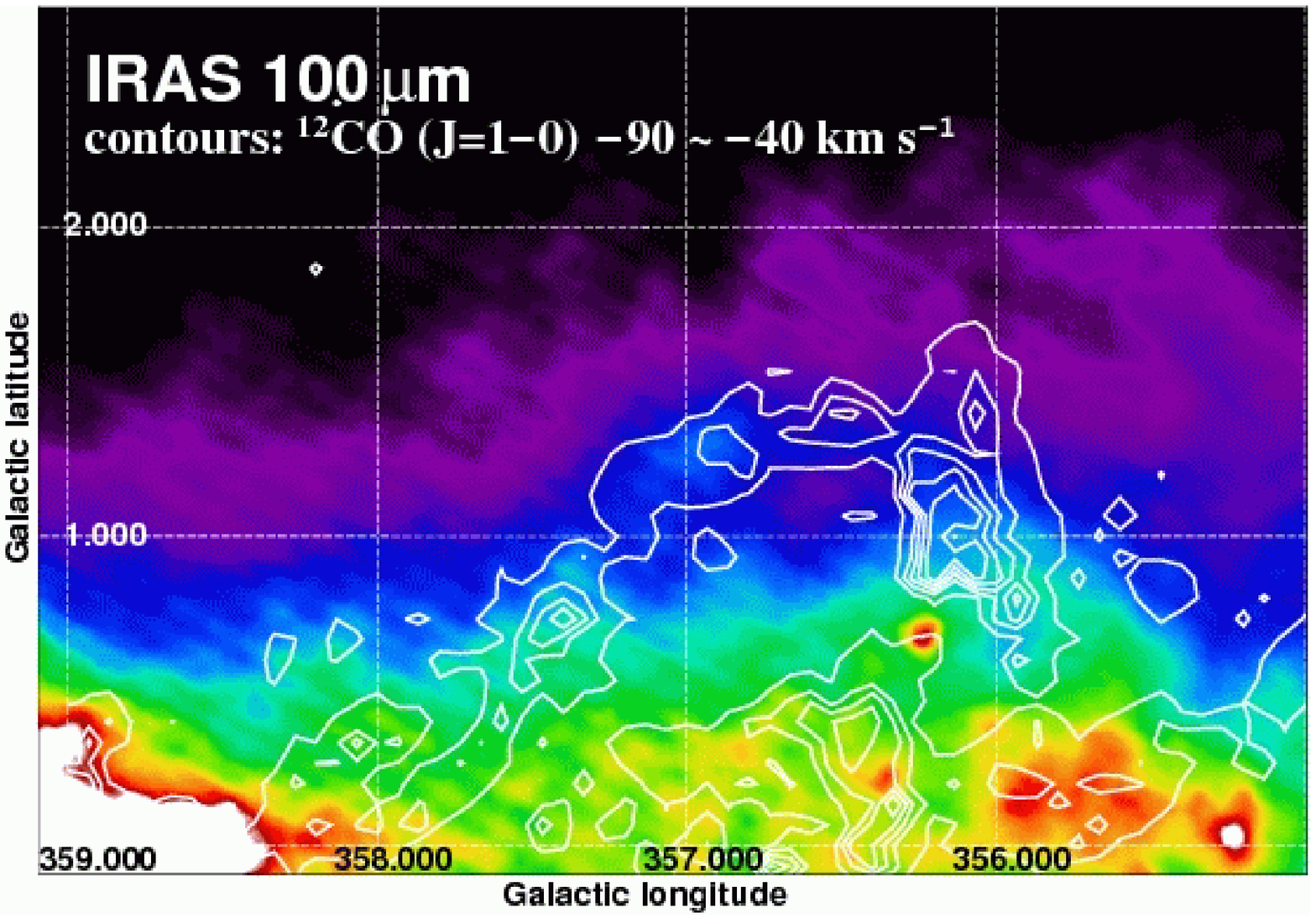}\\
\FigureFile(85mm,85mm){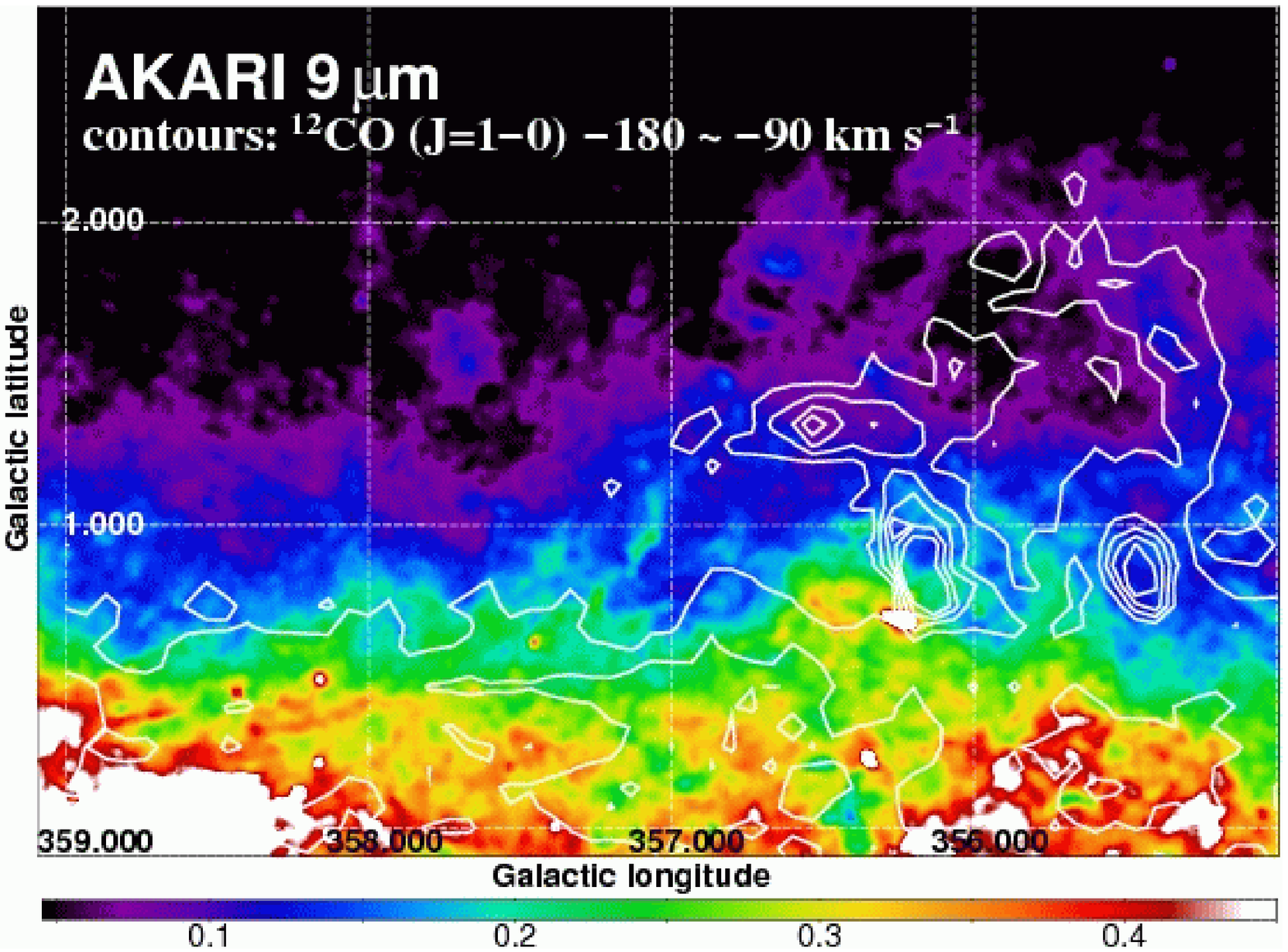}\quad\FigureFile(85mm,85mm){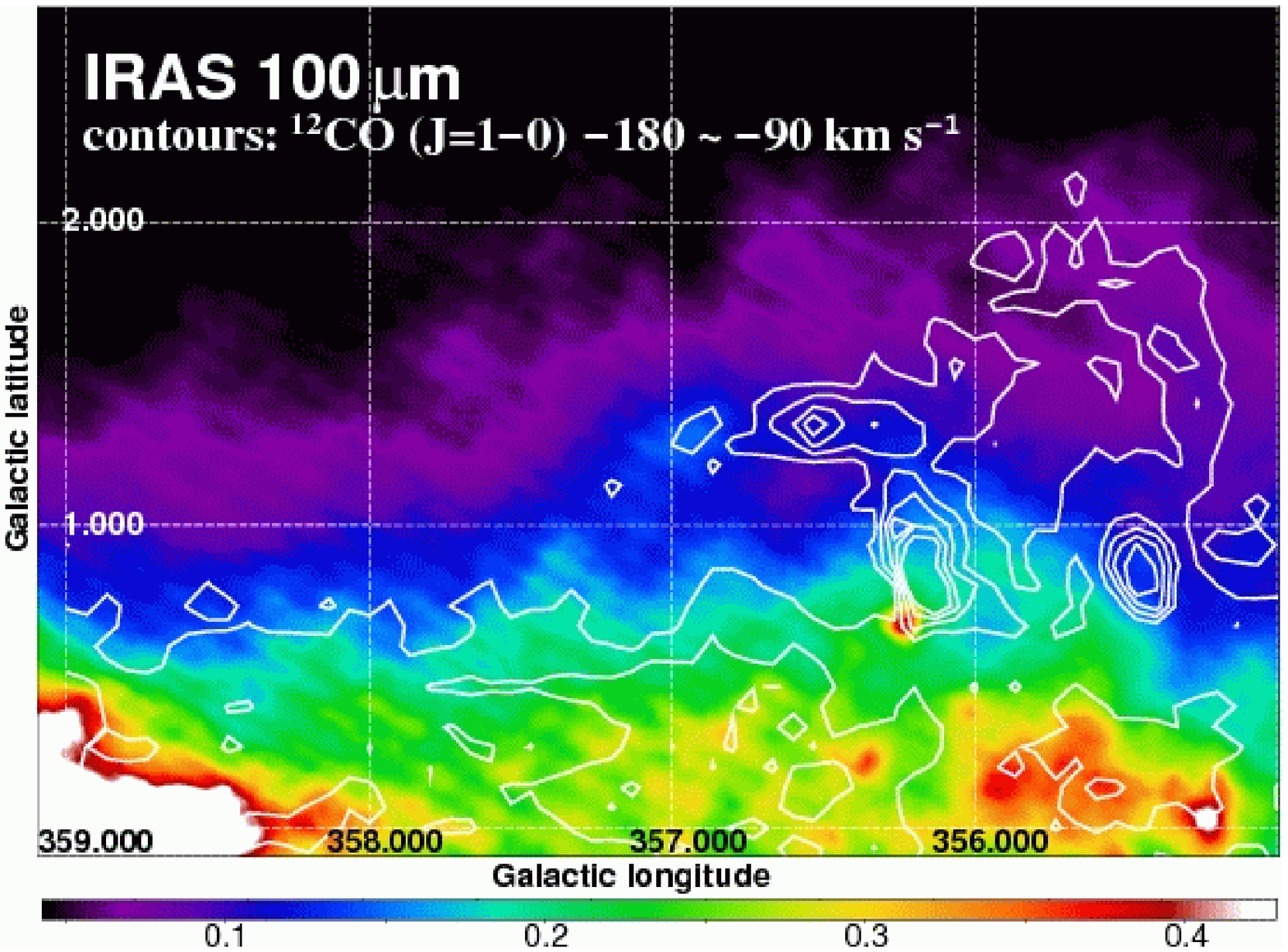}\\
\caption{(Left) AKARI 9 $\mu$m map of an area of about $4^{\circ}\times3^{\circ}$ near the Galactic center ($l=\timeform{355^{\circ}.0}-\timeform{359^{\circ}.1}$ and $b=-\timeform{0^{\circ}.1}-\timeform{2^{\circ}.7}$), together with the contour maps of the NANTEN $^{12}$CO $(J=1-0)$ intensities integrated over the velocity ranges of $-90\sim-40$ km s$^{-1}$ and $-180\sim-90$ km s$^{-1}$, which contain the structures of loops 1 and 2, respectively \citep{Fuk06}. The CO contour levels are linearly spaced between 7 and 90 K km s$^{-1}$ for the former and between 10 and 150 K km s$^{-1}$ for the latter panel. (Right) The same as the left panels, but for the IRAS 100 $\mu$m map of the same area. The color levels are shown at the bottom, given in the fraction of the peak surface brightness of 100 MJy sr$^{-1}$ (preliminary) for the AKARI 9 $\mu$m and 8000 MJy sr$^{-1}$ for the IRAS 100 $\mu$m map.  }
\end{figure}

\begin{figure}
\FigureFile(130mm,130mm){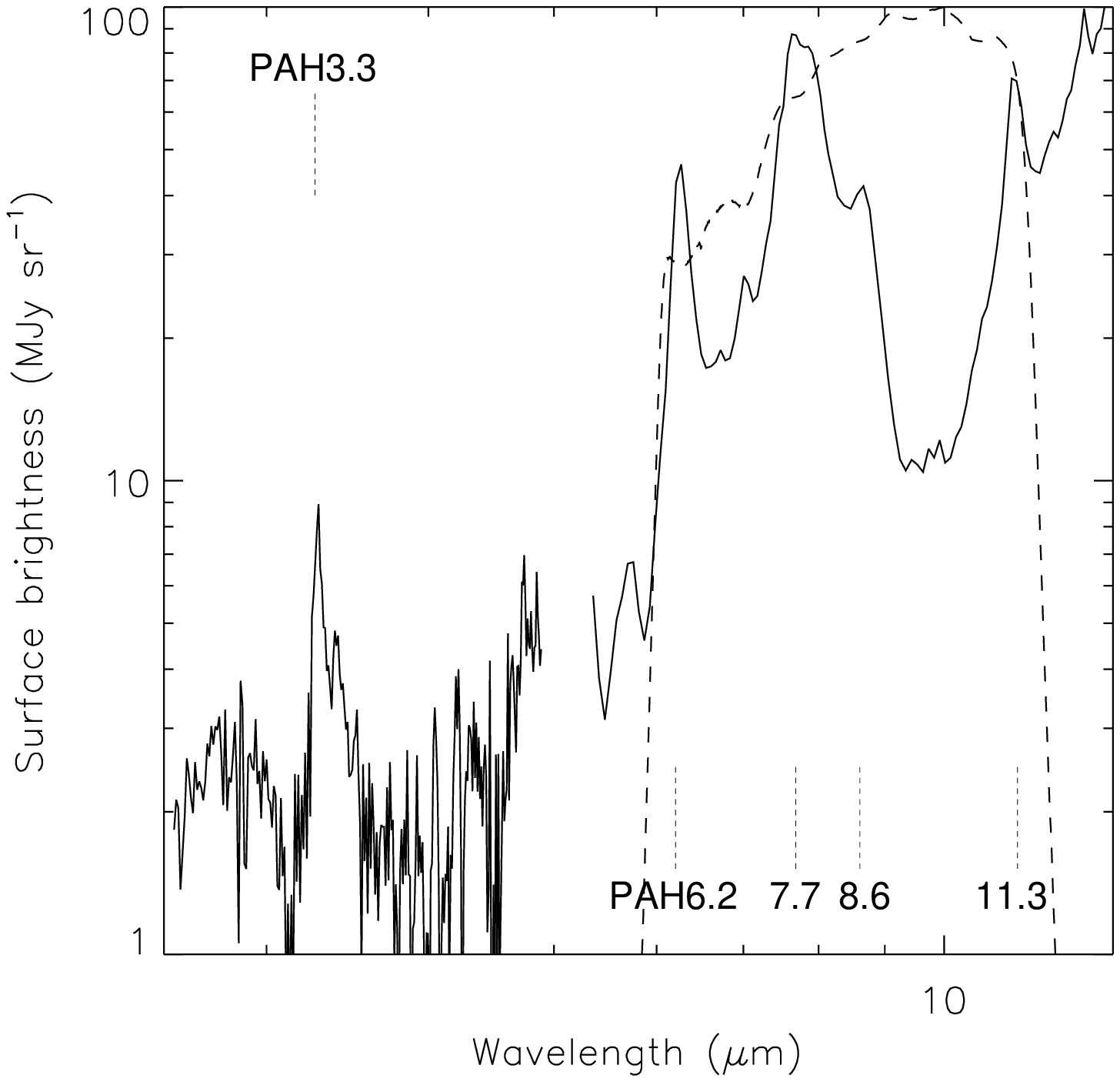}\\
\caption{AKARI 2--14 $\mu$m spectrum of typical Galactic diffuse emission observed at the position of $(l,b)=(\timeform{311^{\circ}.5}, \timeform{0^{\circ}.5})$. The baseline brightness of the mid-IR spectrum is shifted to match continuum levels between the near-IR and the mid-IR spectrum around the wavelength of 5 $\mu$m. The dashed line indicates the relative response curve of the AKARI 9 $\mu$m broad band filter.}
\end{figure}

\section{Results}
\subsection{Mid-infrared 9 $\mu$m diffuse map}

Figure 1 shows the AKARI 9 $\mu$m map of an area of about $4^{\circ}\times3^{\circ}$ near the Galactic center, which is compared with the IRAS 100 $\mu$m map of the same area. As demonstrated by the spectrum in figure 2, the 9 $\mu$m band intensity is mostly attributed to the PAH emissions at wavelengths of 6.2, 7,7, 8.6, and 11.3 $\mu$m. The 100 $\mu$m band flux arises from the thermal emission of submicron size dust grains of astronomical silicate and carbonaceous materials, and thus of a different origin from the 9 $\mu$m band flux. Nevertheless, as can be seen in figure 1, the 9 $\mu$m and the 100 $\mu$m emission show similar spatial distributions. In particular, the emission structure extending from the Galactic plane toward high latitudes is prominent around the area of $l=355^{\circ}-357^{\circ}$ and $b=1^{\circ}-2^{\circ}$ in both bands. Figure 1 also shows the comparisons of the AKARI and IRAS maps with the NANTEN $^{12}$CO $(J=1-0)$ contour maps with the velocity ranges of $-90\sim-40$ km s$^{-1}$ and $-180\sim-90$ km s$^{-1}$, which contain the structures of loops 1 and 2, respectively \citep{Fuk06}. \citet{Tor10b} already pointed out that the IRAS 100 $\mu$m emission is spatially correlated with the CO emission in the loop structure. We here find that the AKARI 9 $\mu$m map also exhibits spatial correspondence to the CO molecular loops. The spatial structure associated with loop 1 is clearer in the IRAS 100 $\mu$m than in the AKARI 9 $\mu$m, while that with loop 2 is clearer in the 9 $\mu$m emission. The top of loop 1 appears to be faint in the PAH emission, making loop 1 less clear in the 9 $\mu$m map. 

\begin{figure}
\FigureFile(130mm,130mm){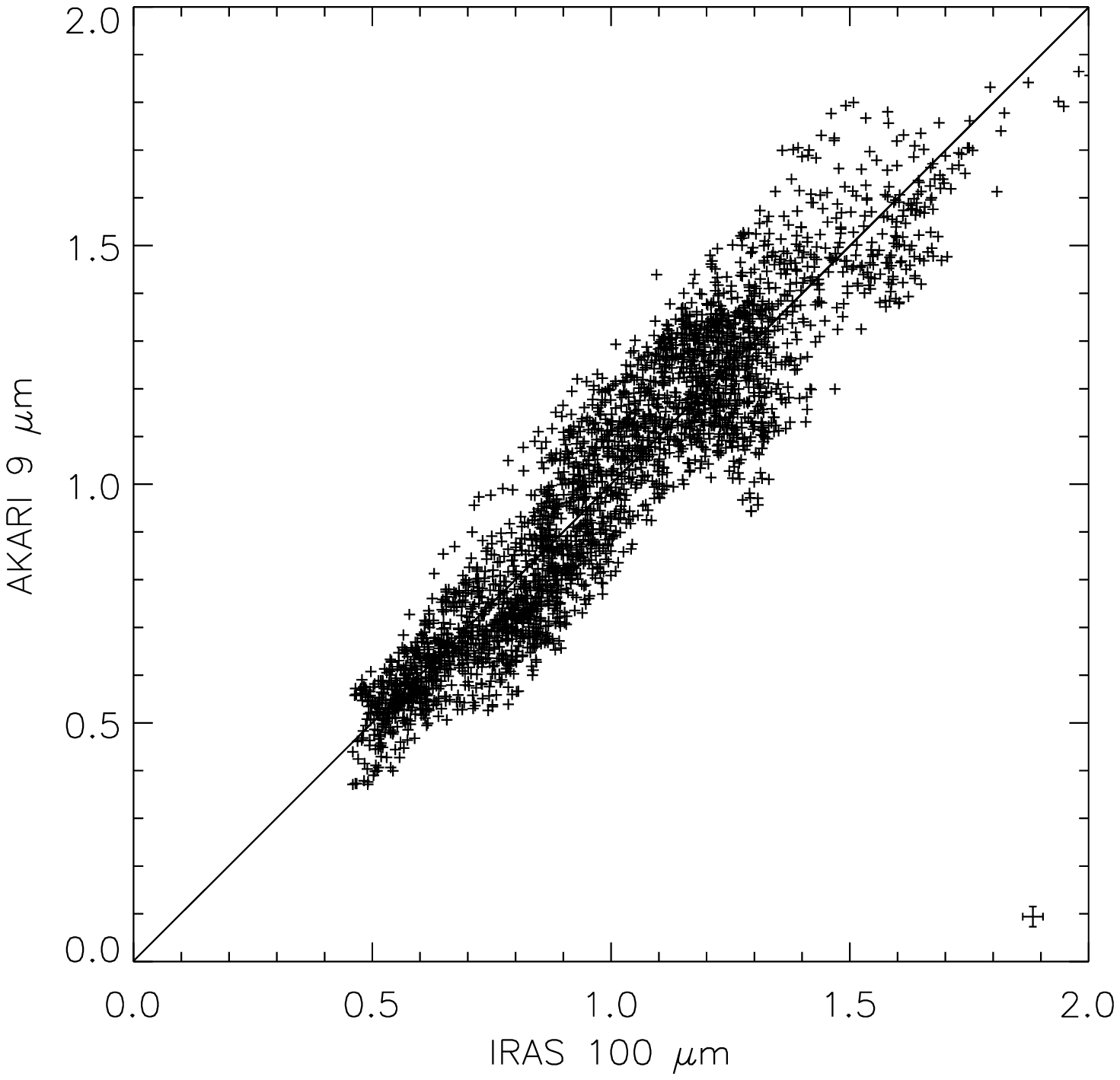}\\
\caption{Correlation plot between the surface brightness of the AKARI 9 $\mu$m and the IRAS 100 $\mu$m map. The data are sampled every 10 spatial bins ($\sim2'$) for an area of $3^{\circ}\times1^{\circ}$ ($l=355^{\circ}-358^{\circ}$ and $b=0^{\circ}-1^{\circ}$). The averaged surface brightness is normalized to unity for each band. Representative error bars of data points near the averages are shown in the lower right-hand corner.}
\end{figure}

\begin{figure}
\FigureFile(85mm,85mm){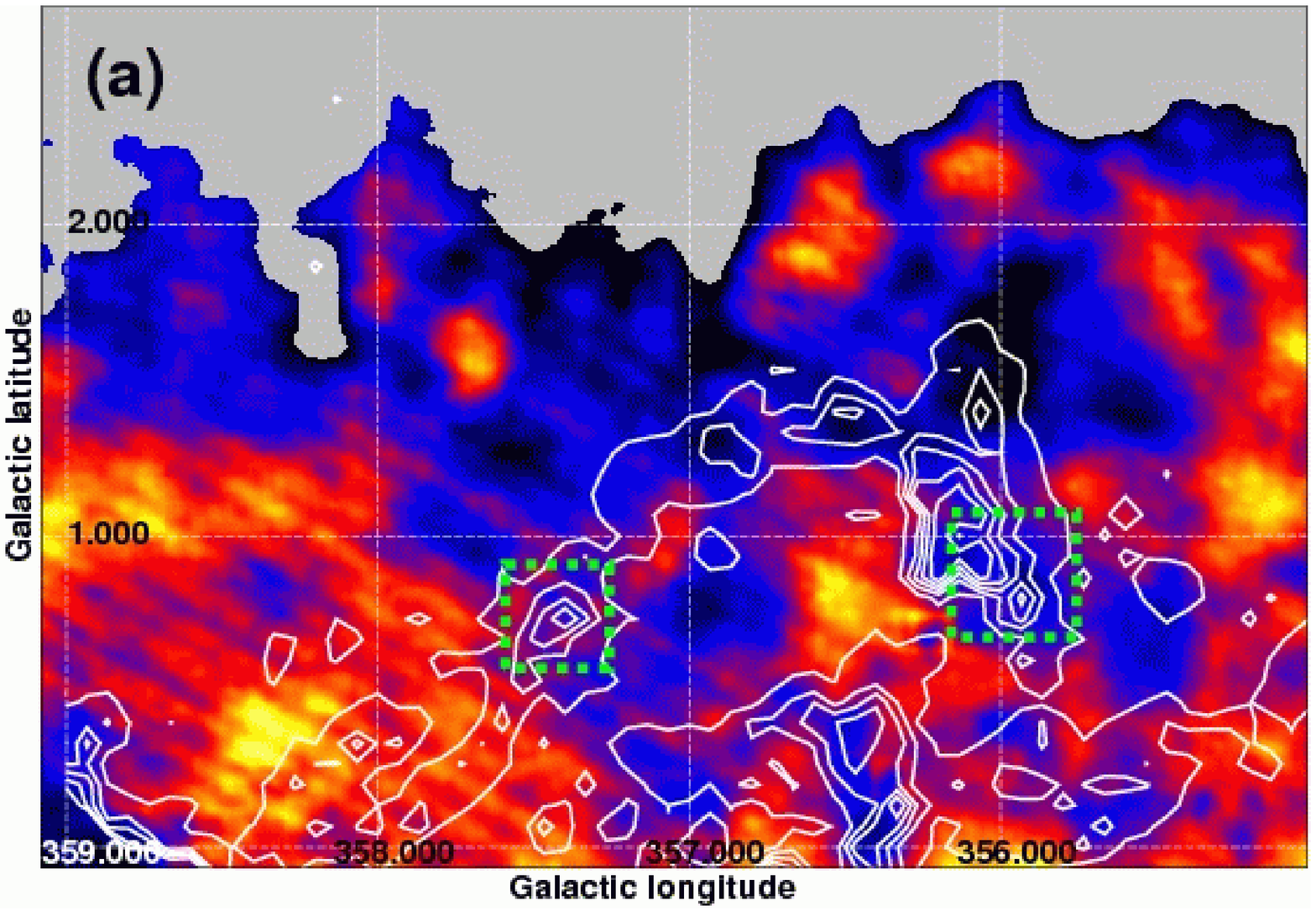}\FigureFile(85mm,85mm){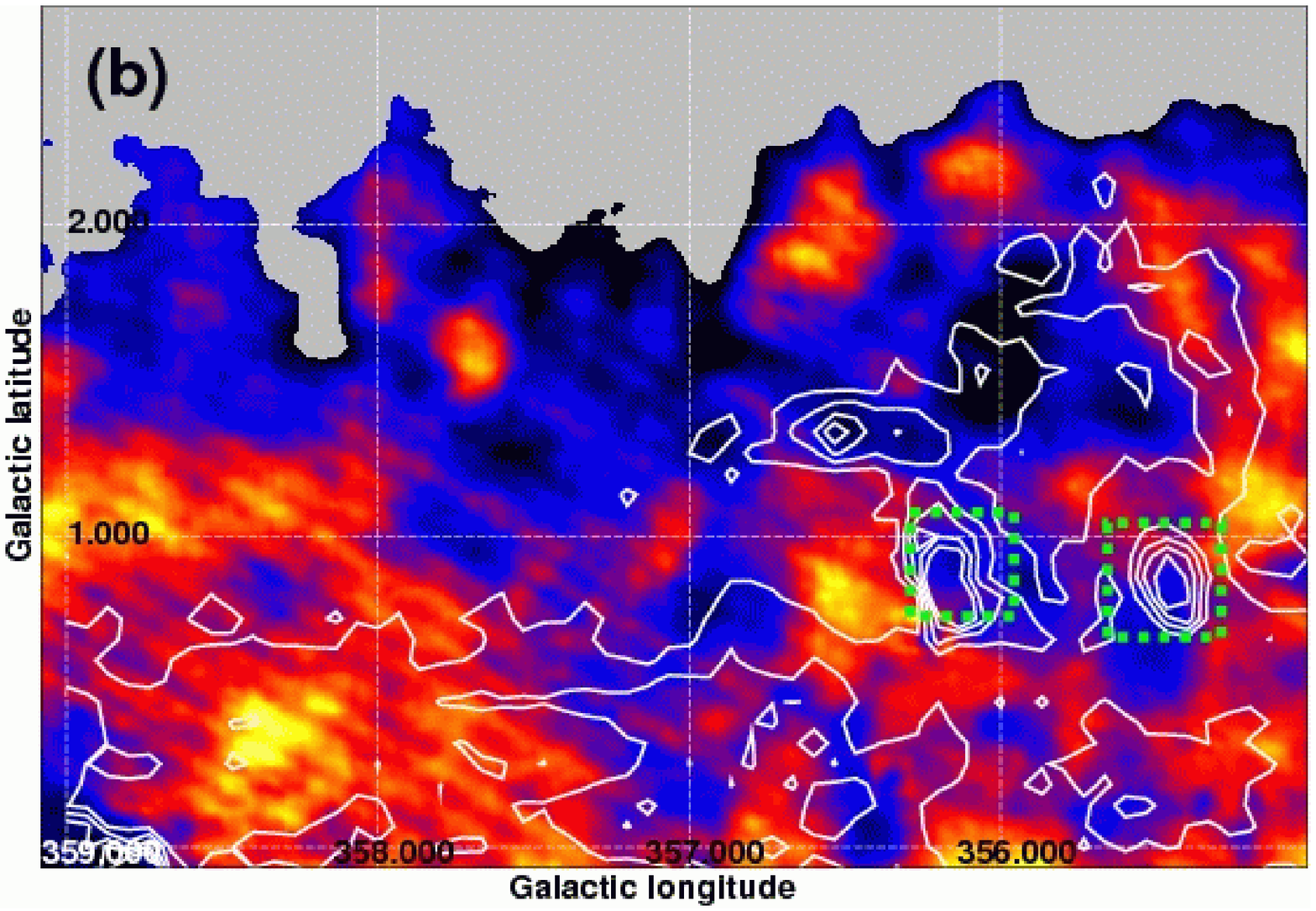}\\
\FigureFile(85mm,85mm){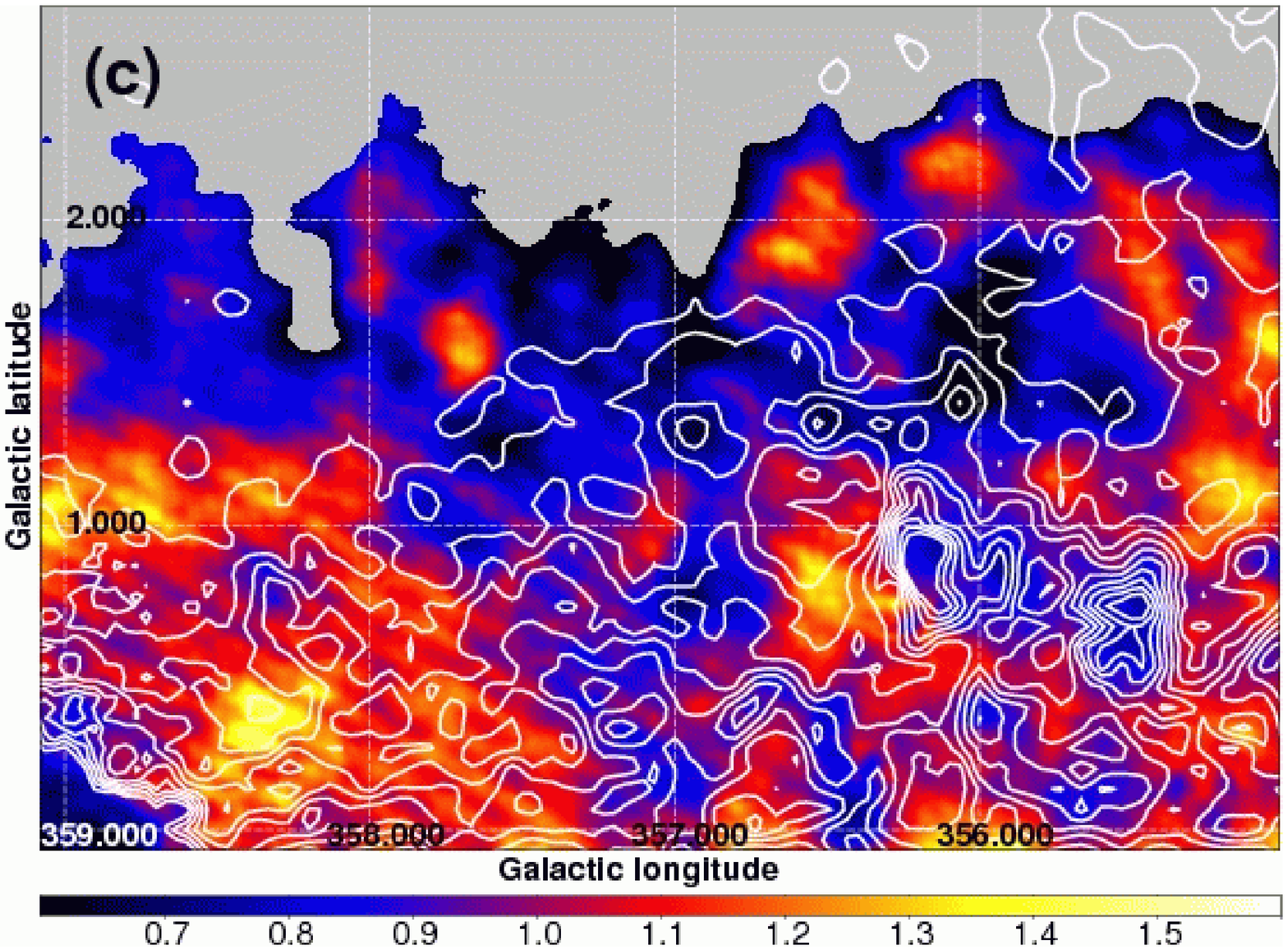}\FigureFile(85mm,85mm){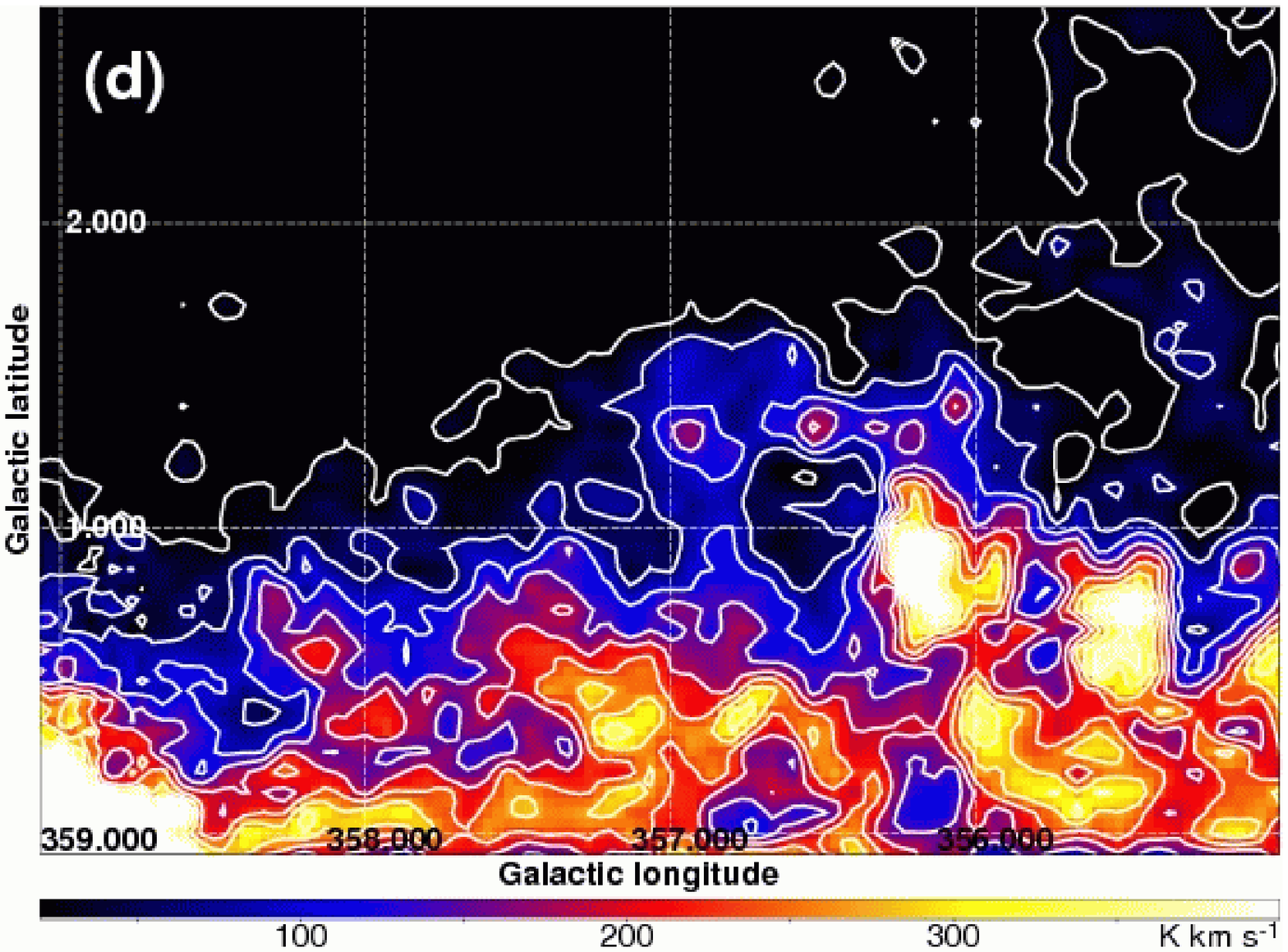}\\
\caption{Distribution of the ratio of the AKARI 9 $\mu$m to the IRAS 100 $\mu$m brightness for the same area as in figure 1, together with the contour maps of the $^{12}$CO $(J=1-0)$ intensities integrated over the velocity ranges of (a) $-90\sim-40$ km s$^{-1}$ (loop 1), (b) $-180\sim-90$ km s$^{-1}$ (loop 2), and (c) $-300\sim200$ km s$^{-1}$. The color levels for the ratio map are shown at the bottom of panel (c); the ratio averaged for an area of $3^{\circ}\times1^{\circ}$ ($l=355^{\circ}-358^{\circ}$ and $b=0^{\circ}-1^{\circ}$) is normalized to unity for each band. Panel (d) shows the color image of the same CO map as in panel (c). The CO contour levels for panels (a) and (b) are the same as in figure 1, while the contour levels for panels (c) and (d) are linearly spaced between 20 and 400 K km s$^{-1}$.  }
\end{figure}

Figure 3 shows the correlation plot between the surface brightness of the AKARI 9 $\mu$m and the IRAS 100 $\mu$m map for an area of $3^{\circ}\times1^{\circ}$ ($l=355^{\circ}-358^{\circ}$ and $b=0^{\circ}-1^{\circ}$) in figure 1, where the averaged surface brightness is normalized to unity for each map. The figure reveals a very tight correlation; the linear-correlation coefficient, $R$, is 0.94 for 2821 data points. This supports that the PAHs are well mixed with the dust grain population as already pointed out by many authors (e.g. Onaka et al. 1996). Yet the figure also exhibits small but significant variations in the relation between the PAH and the dust emission. We examine the spatial distribution of the variations from the ratio map of the 9 $\mu$m to the 100 $\mu$m brightness in figure 4; the spatial resolution of the AKARI 9 $\mu$m map is reduced to the IRAS map resolution of $\sim4'$ before dividing the AKARI map by the IRAS map. The spatial resolution is also comparable to the grid size of the NANTEN $^{12}$CO $(J=1-0)$ map to be compared with the ratio map. The areas with low signal-to-noise ratios are masked, where the surface brightness in the 9 $\mu$m is lower than 7 MJy sr$^{-1}$. As shown in figure 4, the ratios vary within $\pm 30$ \% around the average. It is found that the ratios tend to become lower near the foot points of loops 1 and 2 (indicated by the dashed squares in figures 4a and 4b) than their ambient areas. Figures 4c and 4d shows the NANTEN $^{12}$CO $(J=1-0)$ integrated intensity map with a velocity range of $-300\sim200$ km s$^{-1}$ including both loop structures. It is clear from the figures that the ratios of the 9 $\mu$m to the IRAS 100 $\mu$m brightness show appreciable depression near bright regions in the CO emission.

More quantitatively, figure 5 shows the scatter plot between the NANTEN $^{12}$CO $(J=1-0)$ and the AKARI 9 $\mu$m brightness, both divided by the IRAS 100 $\mu$m brightness, for an area of $3^{\circ}\times1^{\circ}$ ($l=355^{\circ}-358^{\circ}$ and $b=0^{\circ}-1^{\circ}$). 
It is clear from the figure that the data points for the regions near the foot points of loops 1 and 2 (indicated by the red boxes) show relatively low 9 $\mu$m/100 $\mu$m ratios and high CO $(J=1-0)$/100 $\mu$m ratios, producing a significant anti-correlation with $R=-0.58$ for 336 data points. To our knowledge, this is the first result for anti-correlation between PAH and CO emission on such large spatial scales. The 9 $\mu$m/100 $\mu$m ratio most probably reflects the PAH abundance relative to the sub-micron dust grains, revealing the paucity of PAHs at the foot points of loops 1 and 2, as discussed below. 

\begin{figure}
\FigureFile(130mm,130mm){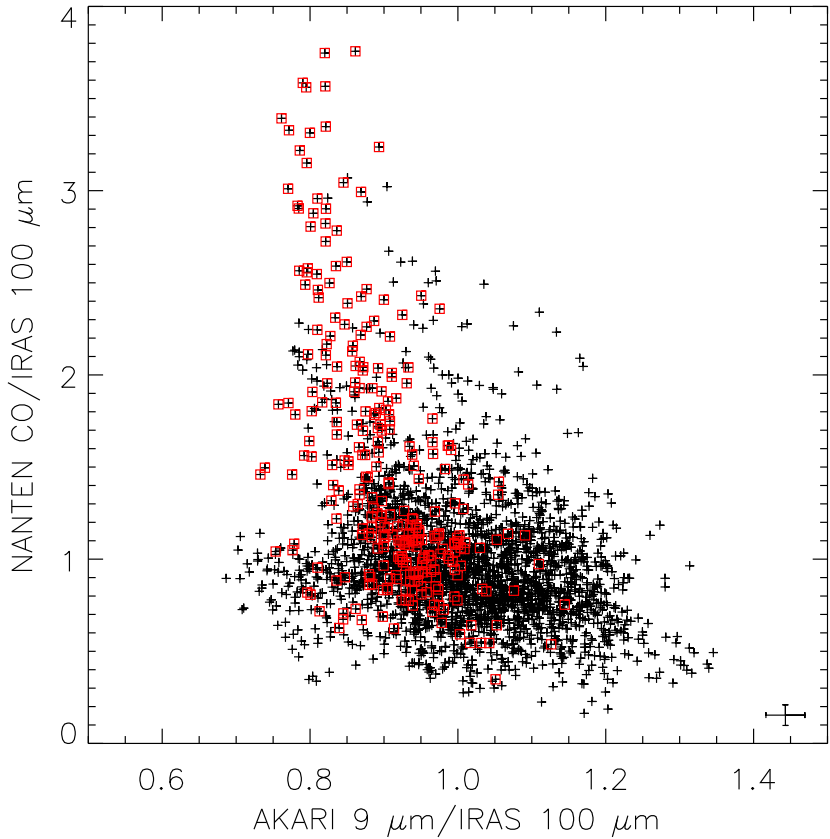}\\
\caption{Scatter plot between the ratio of the NANTEN $^{12}$CO $(J=1-0)$ to the IRAS 100 $\mu$m brightness and that of the AKARI 9 $\mu$m to the IRAS 100 $\mu$m brightness. The $^{12}$CO $(J=1-0)$ intensities are integrated over the velocity range of $-300\sim200$ km s$^{-1}$. The data are sampled every 10 spatial bins ($\sim2'$) for an area of $3^{\circ}\times1^{\circ}$ ($l=355^{\circ}-358^{\circ}$ and $b=0^{\circ}-1^{\circ}$). The brightness averaged for the area is normalized to unity for each. The red boxes indicate the data points obtained from the regions near the foot points of loops 1 and 2, which are indicated by the dashed squares in figure 4. Representative error bars of data points near the averages are shown in the lower right-hand corner.}
\end{figure}

\subsection{Near-infrared $2.5-5$ $\mu$m spectra}

Figure 6 shows the AKARI near-IR 2.6--4.7 $\mu$m spectra. Figures 6a and 6b represent the spectra taken from Galactic on-plane regions near the Galactic center ($l=356^{\circ}-357^{\circ}$; see table 1), while figures 6c and 6d show those from the regions near the foot point and the top of loop 2, respectively, the positions of which are indicated in figure 6e. As compared with the spectrum in figure 2, the resolution of the spectra in figure 6 is considerably degraded due to the data reduction procedure to remove hot pixels from the array images taken in Phase 3. Nevertheless we can recognize the presence of the PAH 3.3 $\mu$m feature and the 3.4--3.6 $\mu$m sub-features as well as the hydrogen recombination Br$\alpha$ line at 4.05 $\mu$m for the Galactic diffuse spectra in figures 6a and 6b. As seen in figures 6c and 6d, however, the loop spectra do not apparently exhibit the 3.3 $\mu$m aromatic hydrocarbon emission, whereas a broad feature appears to exist around $\sim3.5$ $\mu$m. Since the loop spectra were taken on higher-temperature conditions in later Phase 3 (table 1), care must be taken about the variability of dark levels, which is not reflected in the errors given by the pipeline process. The solid curves in figures 6c and 6d illustrate how the spectra change with $\pm$5 ADU offsets for non-dispersive dark levels. As can be seen in the figure, the broad feature becomes weak (figure 6c) or disappears (figure 6d) for the minus offset, because the spectral response curve has a broad structure around 3.5 $\mu$m.

\begin{figure}
\begin{minipage}[]{97mm}
\FigureFile(48mm,48mm){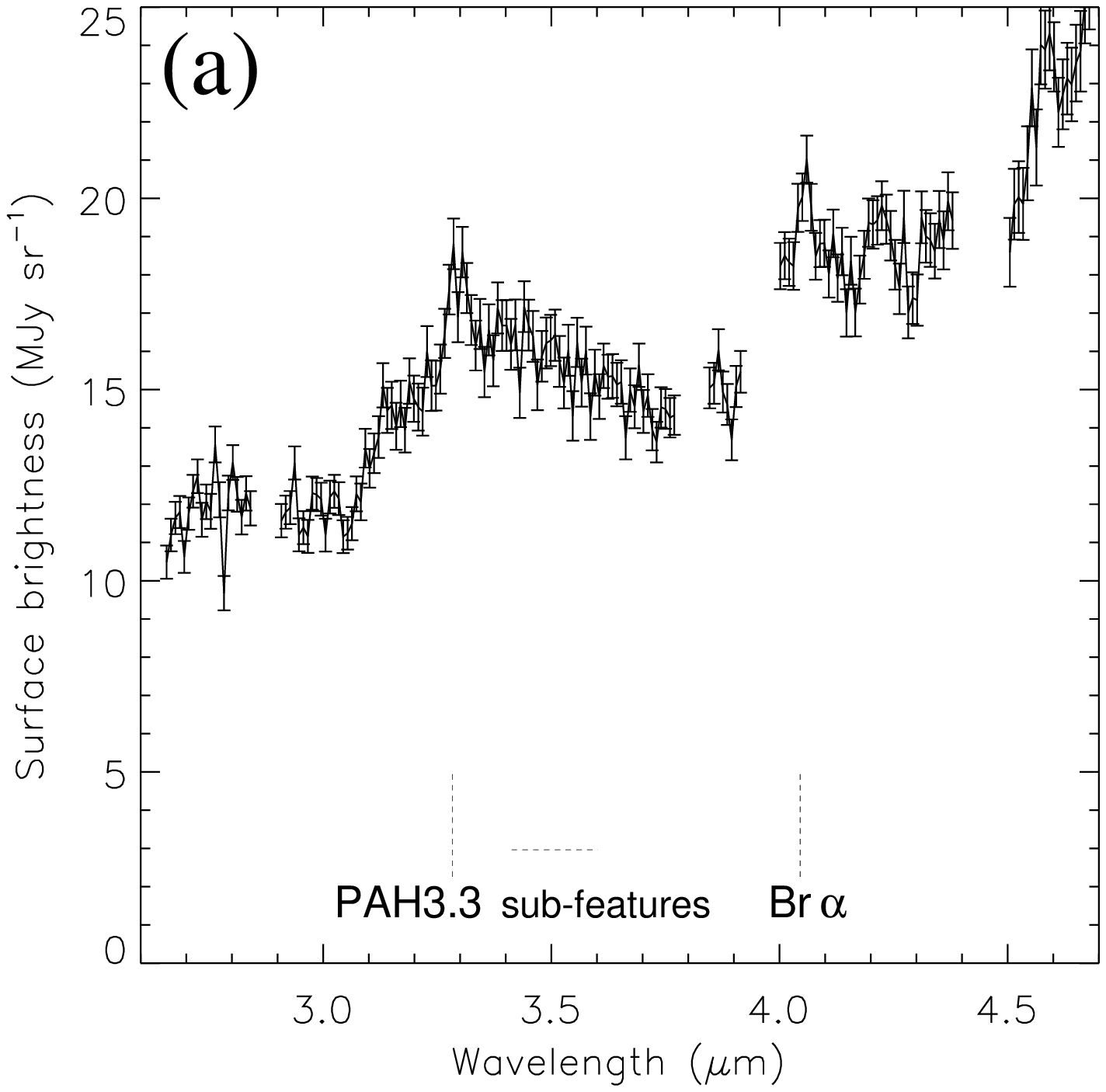}\FigureFile(48mm,48mm){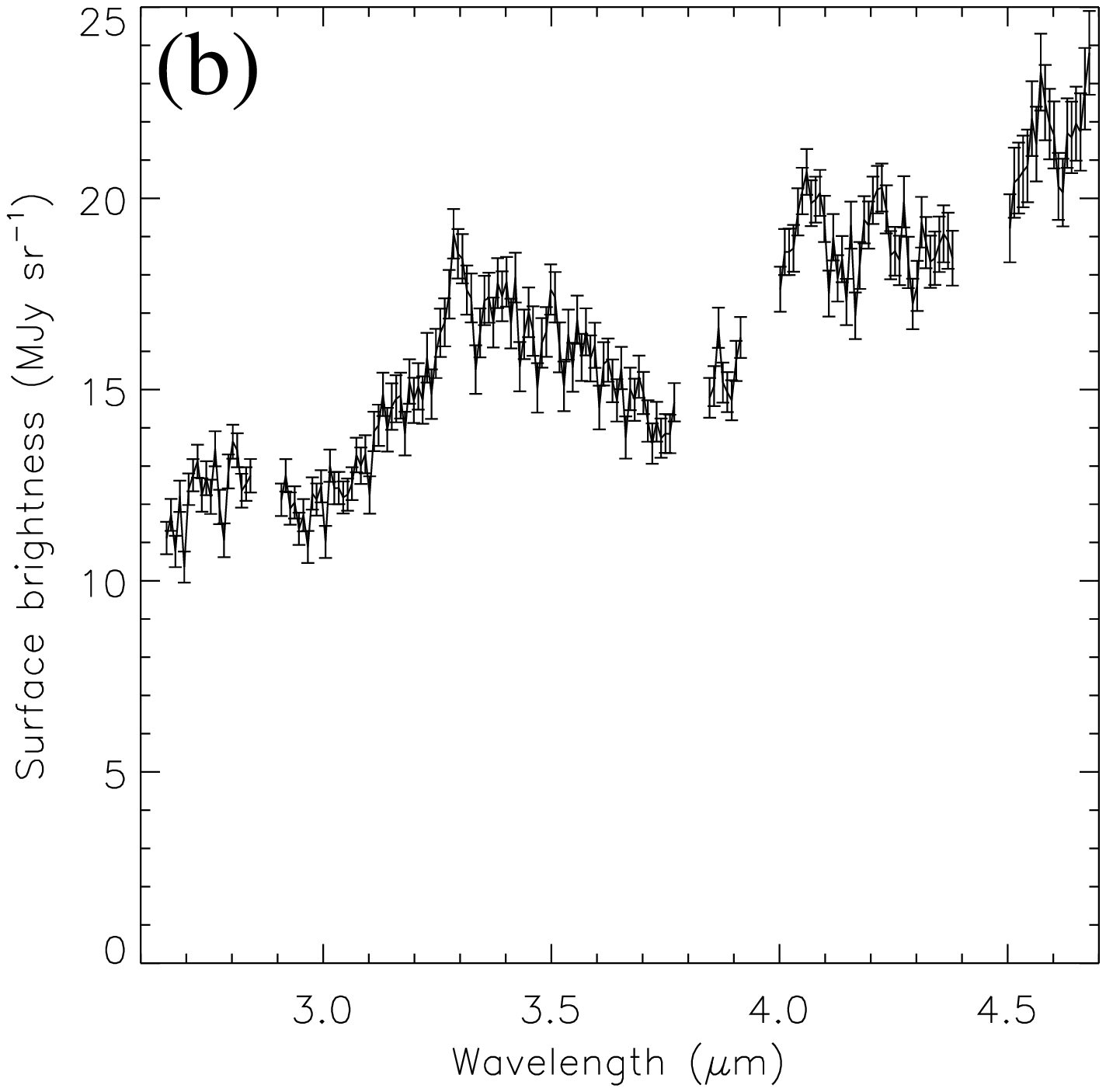}\\
\FigureFile(48mm,48mm){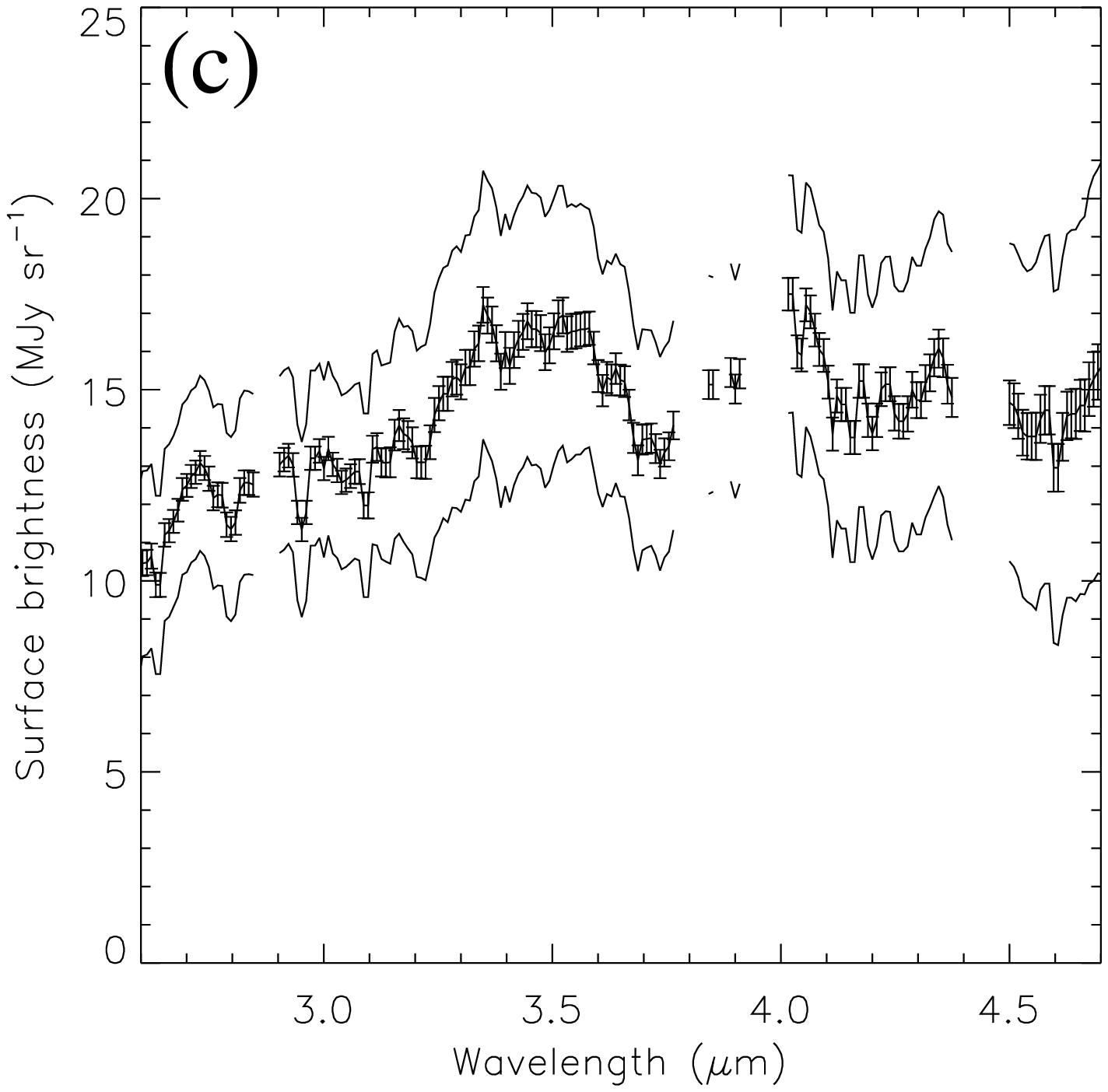}\FigureFile(48mm,48mm){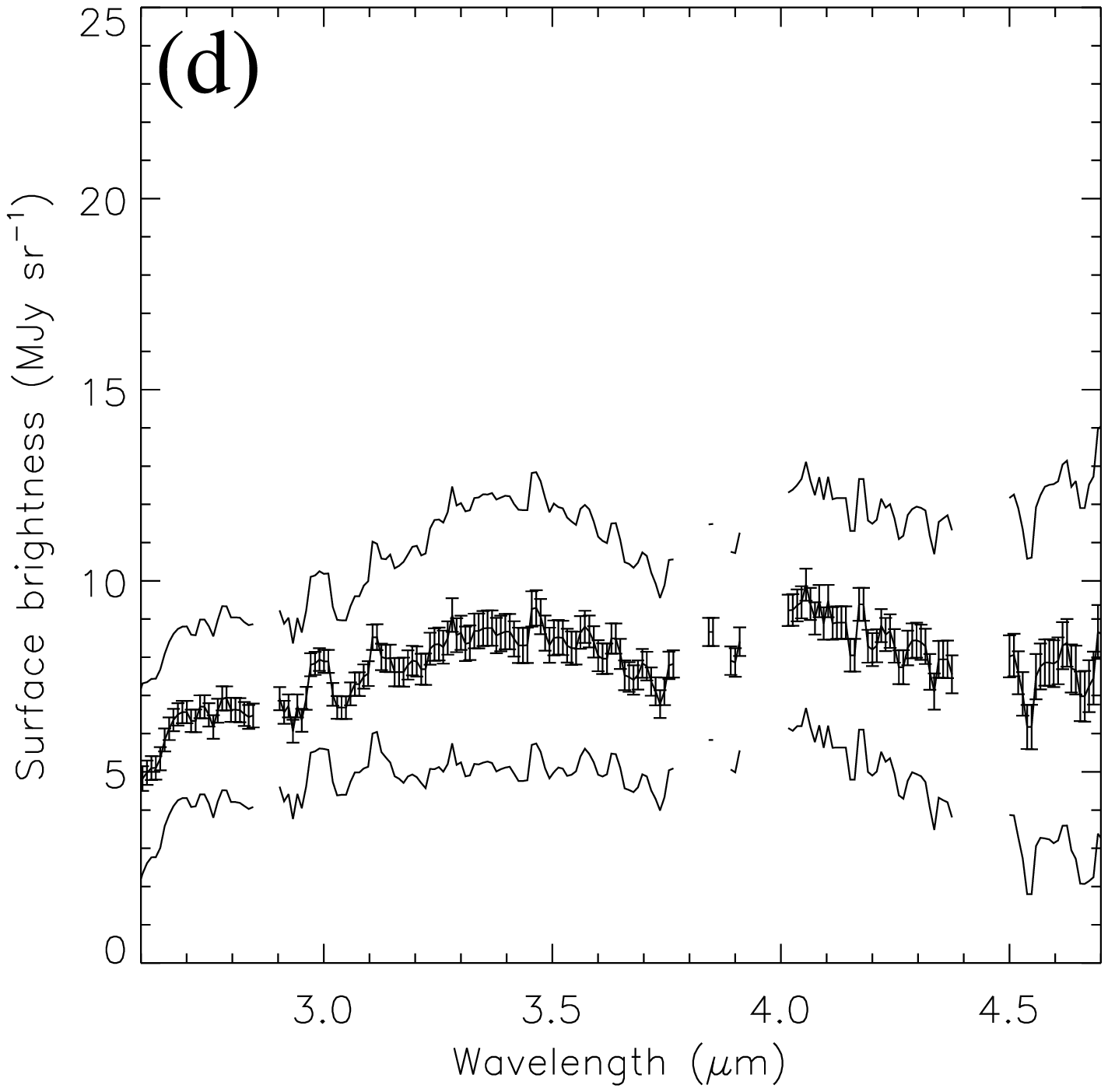}
\end{minipage}
\begin{minipage}[]{71mm}
\FigureFile(71mm,71mm){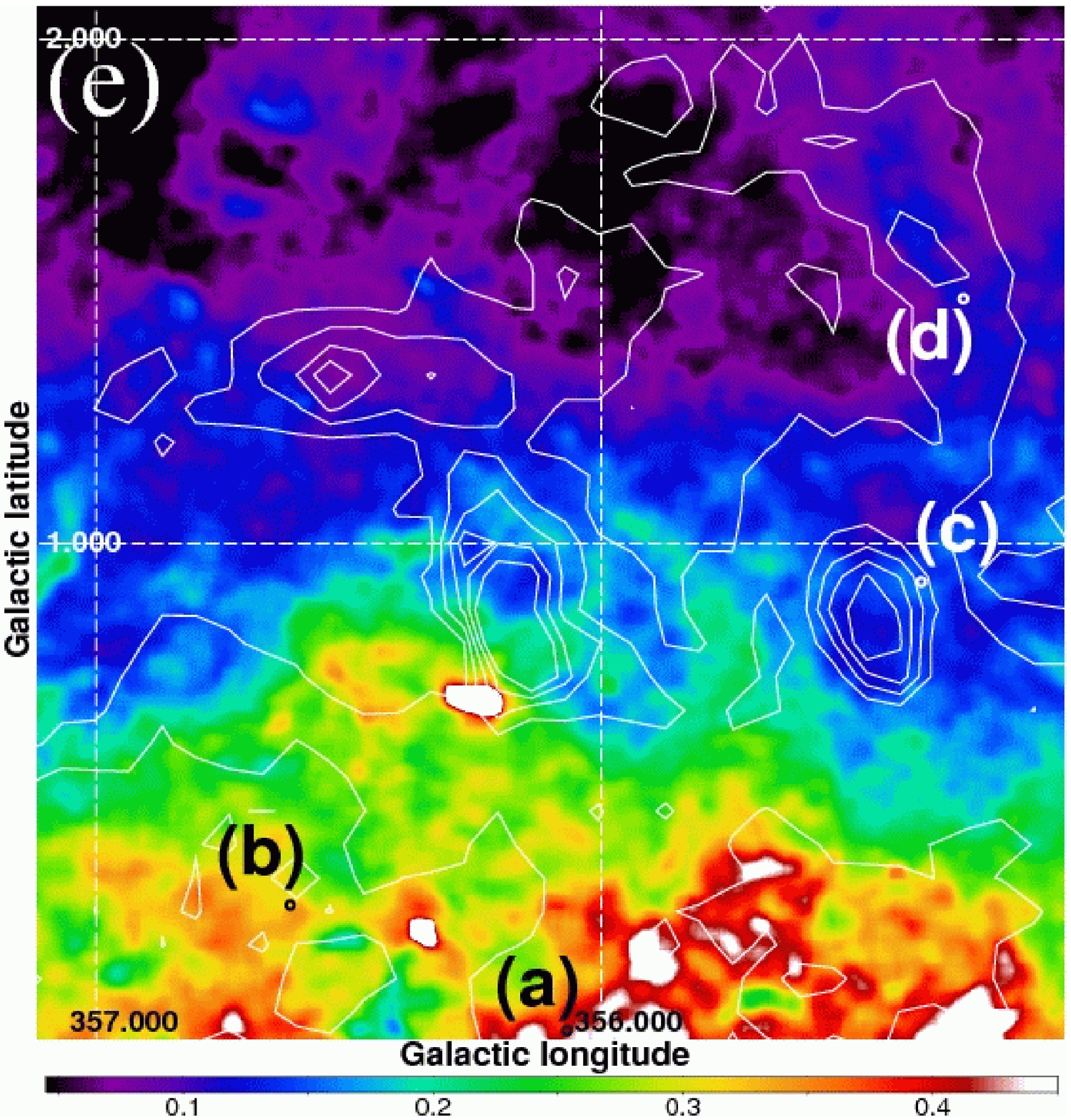}
\end{minipage}
\caption{AKARI near-IR 2.6--4.7 $\mu$m spectra taken from Galactic on-plane regions near the Galactic center ($l=356^{\circ}-357^{\circ}$; panels (a) and (b)) and from the regions near the foot point and the top of loop 2 (panels (c) and (d), respectively). Instrumental artifacts commonly seen among the spectra are masked out. The solid curves in panels (c) and (d) illustrate how the spectra change with $\pm$5 ADU offsets for the dark levels (see text for details). The positions of the centers of the sub-slit apertures used to derive the spectra are indicated in panel (e), which is the close-up map of the lower left-hand panel of figure 1 around loop 2.  }
\end{figure}

\begin{figure}
\FigureFile(48mm,48mm){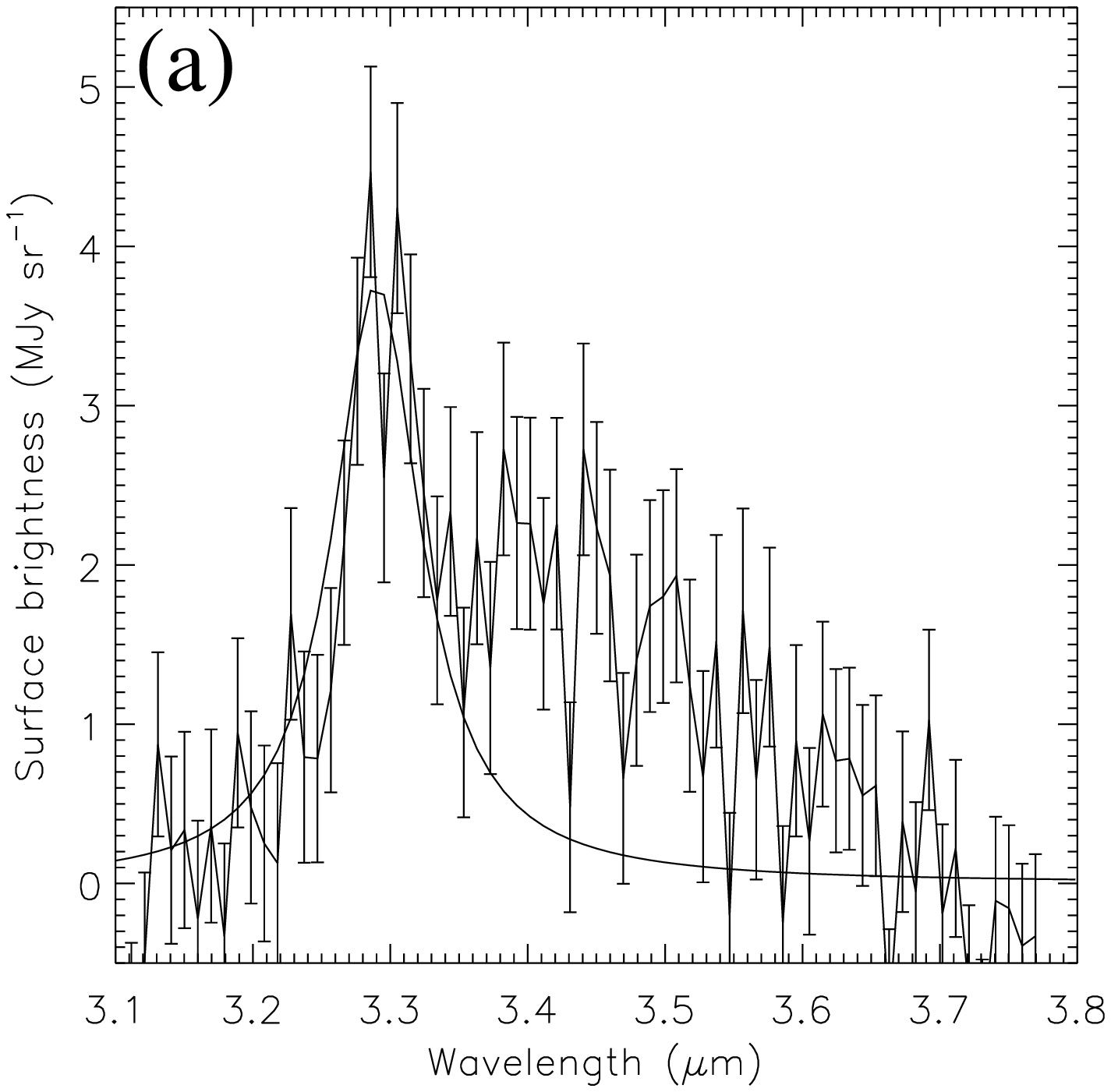}\quad\FigureFile(48mm,48mm){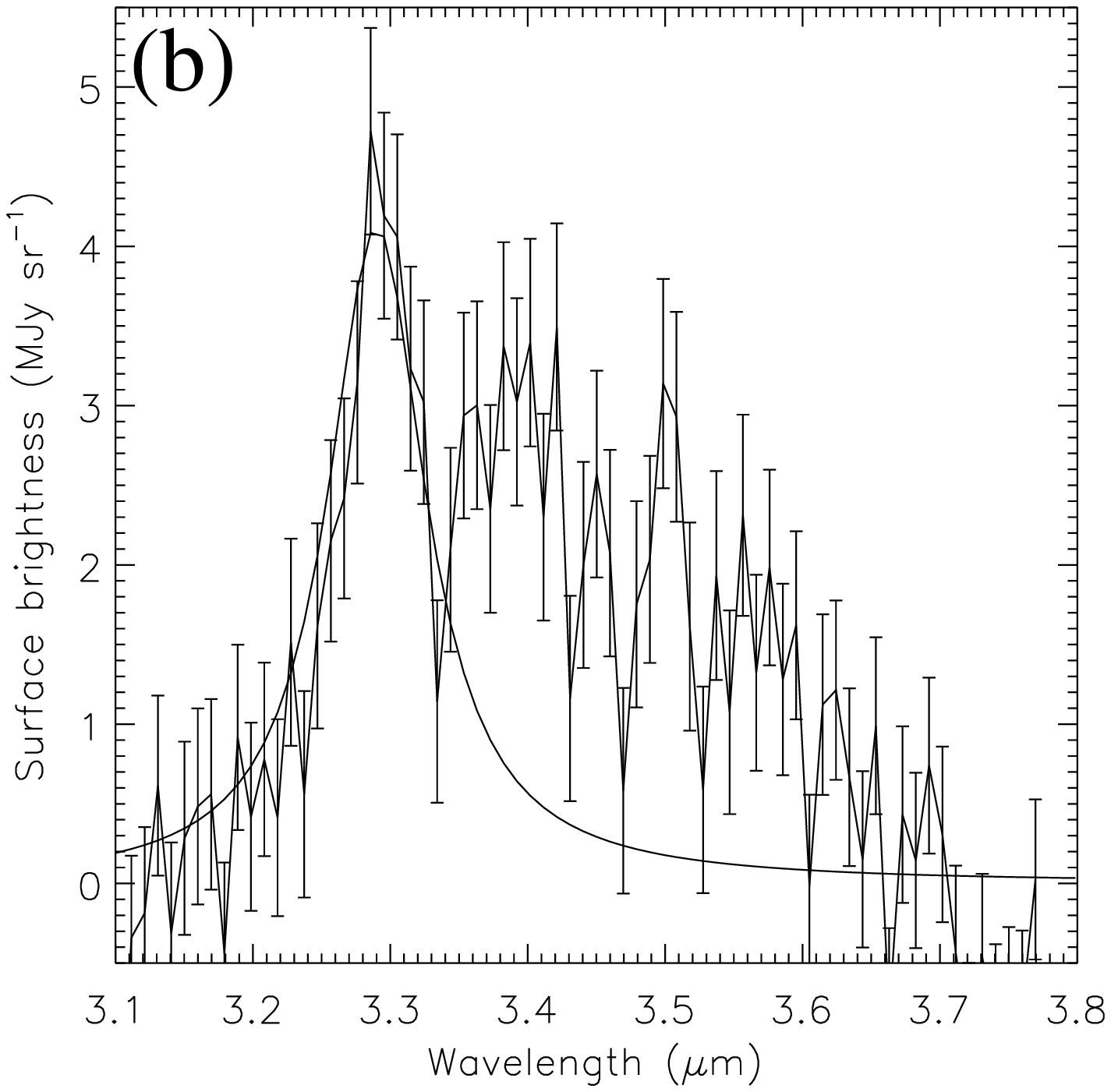}\\
\FigureFile(48mm,48mm){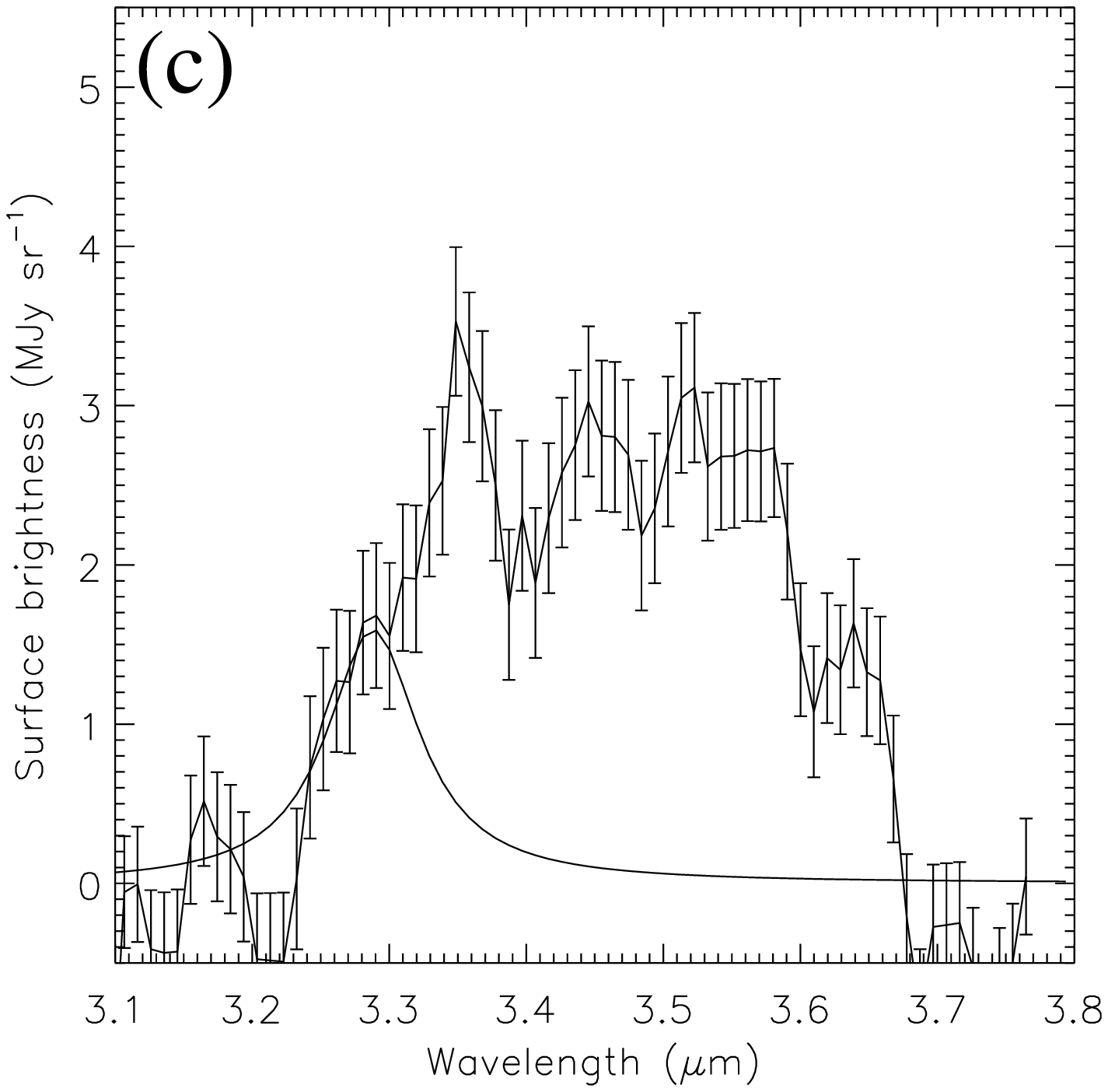}\quad\FigureFile(48mm,48mm){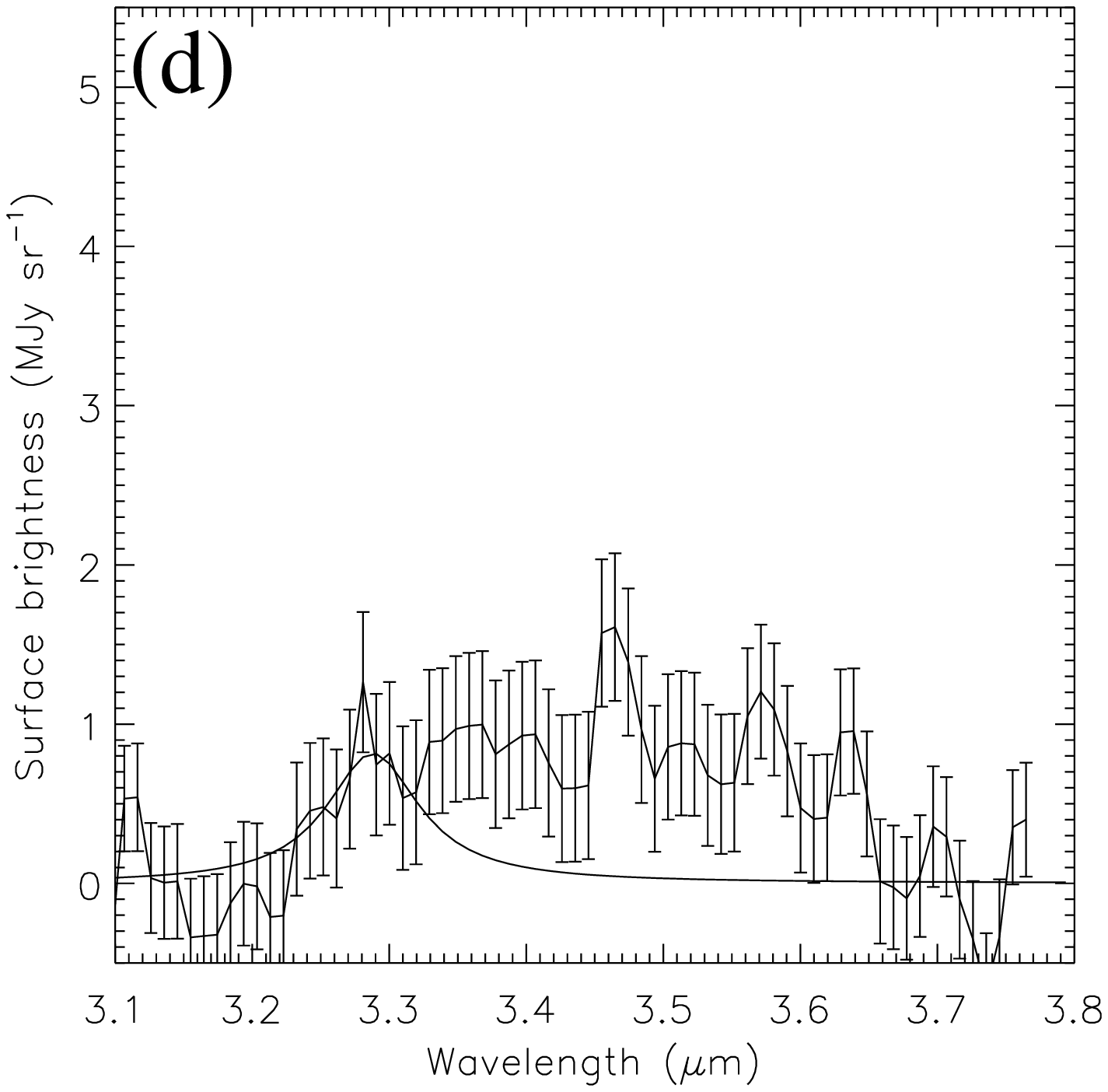}\quad\FigureFile(48mm,48mm){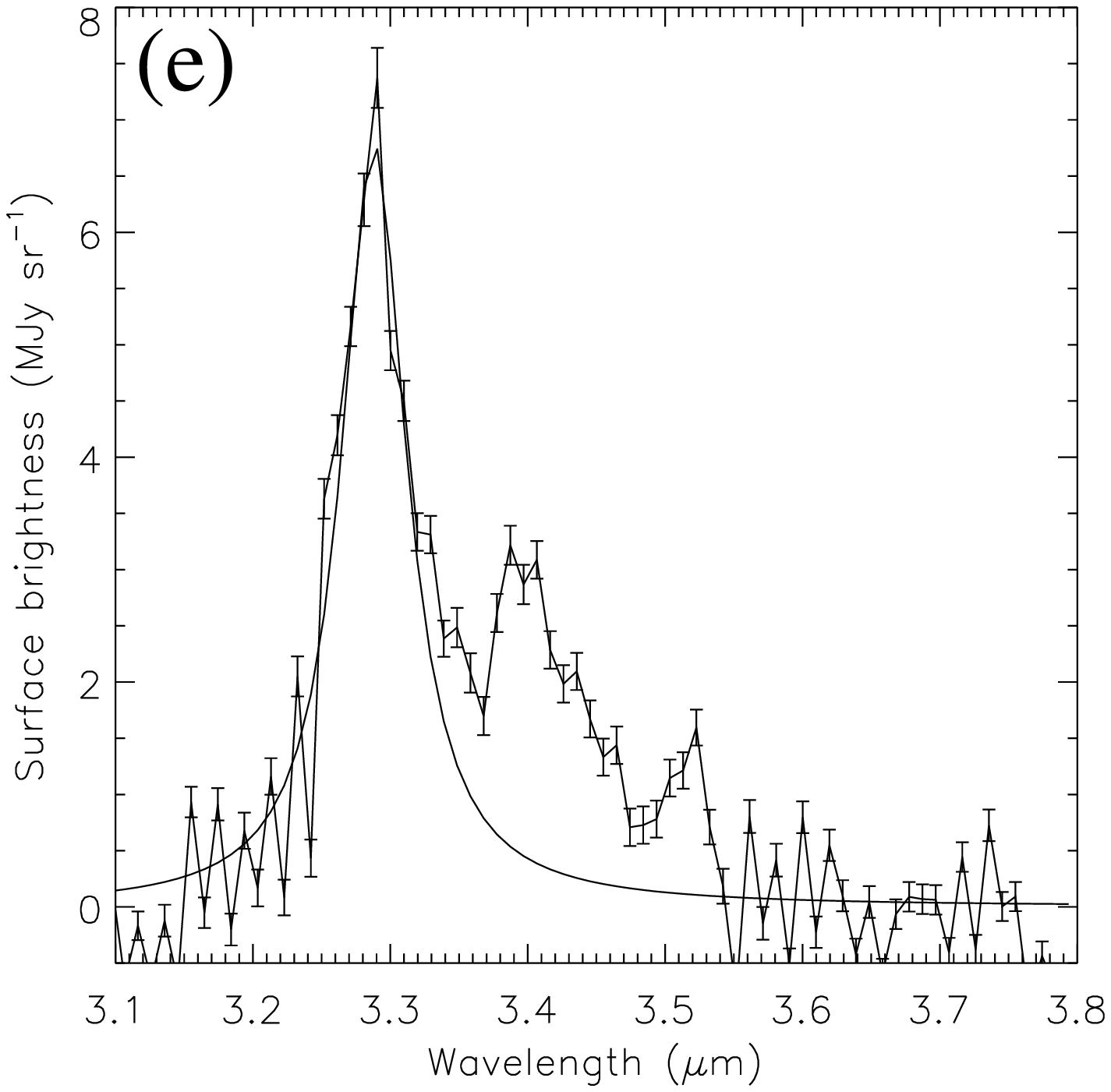}\\
\caption{Continuum-subtracted 3.1--3.8 $\mu$m spectra, where the linear baseline is determined by the wavelength ranges of 3.1--3.2 $\mu$m and 3.65--3.75 $\mu$m and subtracted for each spectrum. Panels (a)--(d) are from the spectra in figures 6a--d, respectively, while panel (e) is from figure 2. 
The solid curves are the best-fit Drude profiles representing the contribution of the PAH 3.3 $\mu$m emission. Considering the difference in the spectral resolution and the feature profile, slightly different fitting ranges are adopted: 3.2--3.35 $\mu$m for panels (a) and (b), 3.2--3.31 $\mu$m for panels (c) and (d), and 3.2--3.32 $\mu$m for panel (e).}
\end{figure}

To clarify the spectral features at wavelengths of 3.2--3.6 $\mu$m, we subtract an underlying continuum from each spectrum in figures 2 and 6 by approximating the continuum by a linear baseline, which is determined by the spectral ranges of 3.1--3.2 $\mu$m and 3.65--3.75 $\mu$m. Figure 7 shows the baseline-subtracted 3.1--3.8 $\mu$m spectra. 
We fit a Drude profile to the spectra in figure 7 to visualize the contribution of the PAH 3.3 $\mu$m feature \citep{Smi07}. The central wavelength of the profile is fixed to be 3.29 $\mu$m in every case \citep{All89}.
The fitting result clearly demonstrates that the spectra of on-plane regions exhibit the usually strong PAH 3.3 $\mu$m emission accompanied by the 3.4--3.6 $\mu$m sub-features, whereas the spectra near the foot point and the top of loop 2 do not require the presence of the PAH 3.3 $\mu$m emission. Although the broad feature at 3.4--3.6 $\mu$m can potentially include larger uncertainties as described above, the 3.3 $\mu$m narrow feature would not be affected by such systematic errors. 

\section{Discussion}
In figures 4 and 5, we find the trend that the ratio of the 9 $\mu$m to the 100 $\mu$m brightness becomes lower as the CO emission is brighter. Larger interstellar extinction in the mid-IR than in the far-IR systematically lowers the ratios of 9 $\mu$m to 100 $\mu$m brightness in dense gas regions. The dominance of neutral PAHs over ionized ones in very dense gas regions could also lower the ratios, because neutral PAHs emit much less in the 6--9 $\mu$m region than ionized PAHs \citep{All99, Pee02, Dra07}. 
However, in figure 8, we find that there is no significant correlation between the ratio of the 9 $\mu$m to the 100 $\mu$m brightness and the 100 $\mu$m brightness. Because the 100 $\mu$m brightness should trace the total gas density, the absence of the correlation in the figure demonstrates that the 9 $\mu$m to the 100 $\mu$m ratio is not dependent upon gas density or extinction effects.

The ratio of the CO to the dust emission in figure 5 is a quantity not sensitive to the gas density but rather to the gas temperature. 
Indeed unusually high gas temperatures (30--100 K) are observed in the regions associated with the foot points of loops 1 and 2, where there are no radiative heating sources \citep{Tor10a, Kud11}. Thus the high gas temperature, not high gas density, is likely related to the lower ratios of 9 $\mu$m to 100 $\mu$m brightness, suggesting the destruction of PAHs in the regions where the shock heating of gas takes place. 
More precisely, the observed increase in the ratio of the CO to the dust emission can also be explained by increase in the ratio of molecular gas to total gas or the gas-to-dust ratio. However, because the 100 $\mu$m brightness (i.e. total gas density) does not apparently increase at the foot points (figures 1 and 8), the former would call for decreases in H{\small I} density to cancel out increases in molecular gas density toward the foot points, which are not observed \citep{Tor10b}. The latter would suggest decrease in metallicity by a factor of $\sim 5$ over the local areas, which is very unlikely. 

As seen in figures 5 and 8, the suppression of the 9 $\mu$m emission relative to the 100 $\mu$m is only $\sim$20 \% from the average. 
However, we have to consider a line-of-sight projection effect for the IR emission; for example, the eastern foot point of loop 1 shows a relatively weak suppression (figure 4b), presumably due to the presence of unassociated clouds lying at lower Galactic latitudes. In other words, the above level of the suppression implies the efficient destruction of PAHs in local areas. Therefore the result calls for such a condition in the magnetic flotation picture that the falling speed of the gas along the loop should be as fast as 50--100 km s$^{-1}$ \citep{Mic10a}. In fact, in the CO emission, the foot points show large velocity dispersions of 40--80 km s$^{-1}$ \citep{Fuk06}, and hence the above shock speed is reasonable. 
Figure 4c further suggests the presence of local regions other than the foot points of loops 1 and 2, which similarly show the suppression of the 9 $\mu$m emission relative to the 100 $\mu$m emission. One of such regions corresponds to the western foot point of loop 3 around $(l,b)=(\timeform{355.6^{\circ}}, \timeform{0.7^{\circ}})$, another molecular loop found with the NANTEN telescope by \citet{Fuj09}. Possible presence of other molecular-loop-like structures are to be investigated in the CO emission.
 
\begin{figure}
\FigureFile(130mm,130mm){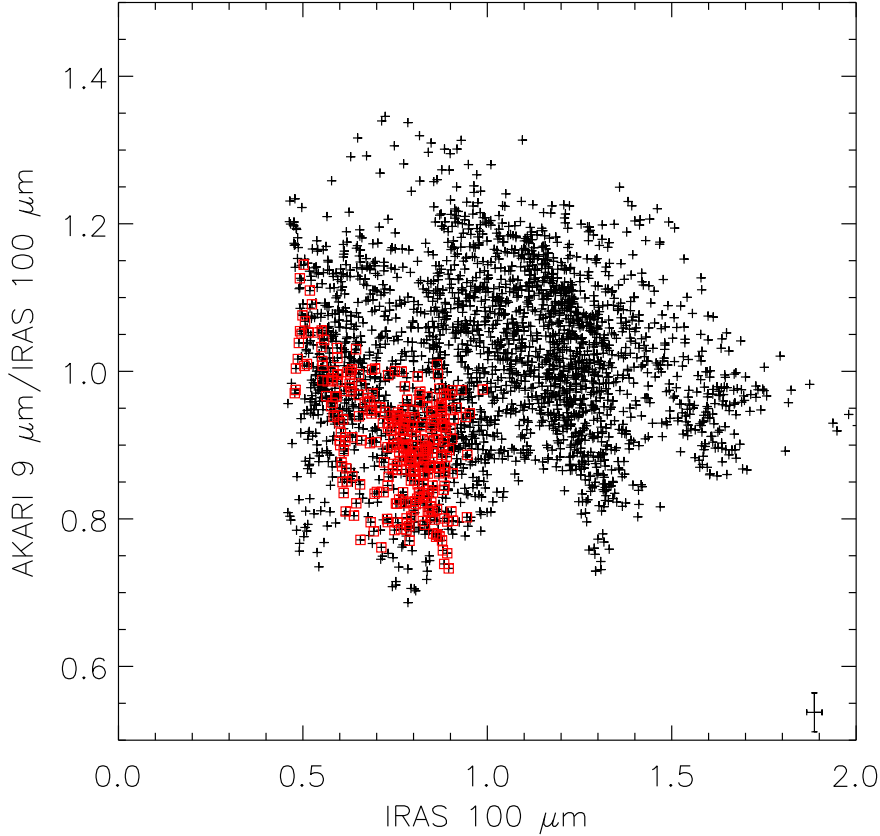}\\
\caption{Scatter plot between the ratio of the AKARI 9 $\mu$m to the IRAS 100 $\mu$m brightness and the IRAS 100 $\mu$m brightness. The data sample is the same as in figure 5. }
\end{figure}

\begin{figure}
\FigureFile(130mm,130mm){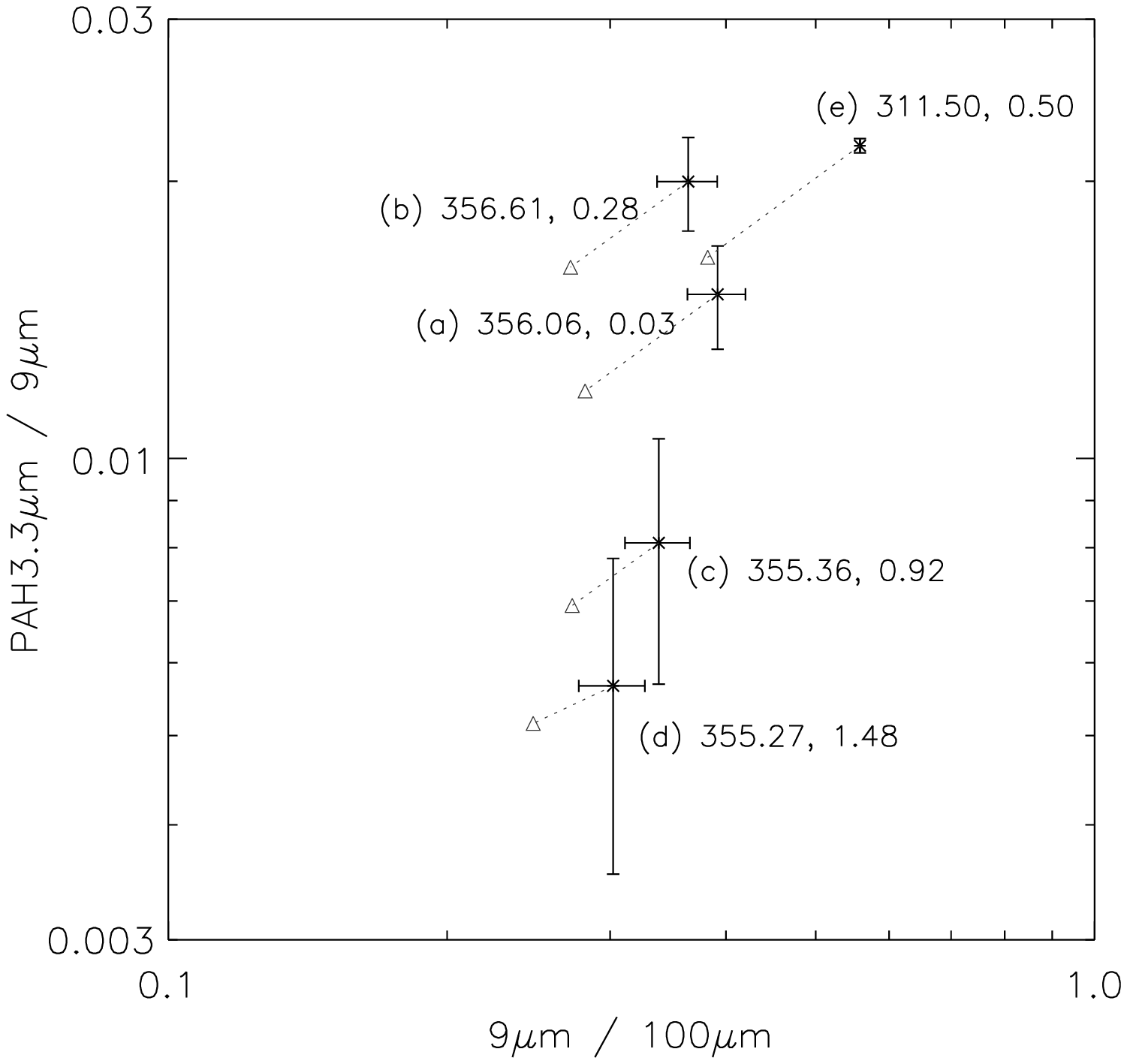}\\
\caption{Ratio of the PAH 3.3 $\mu$m to the 9 $\mu$m band intensity plotted against the ratio of the 9 $\mu$m to the 100 $\mu$m band intensity. The intensities with the error bars are those integrated over the corresponding frequency ranges and corrected for interstellar extinction (see text). The triangles indicate the intensities not corrected for the extinction. The alphabets and the pairs of the numbers attached to the data points denote the spectra in figures 7a--e and their positions in the Galactic coordinates, respectively. 
}
\end{figure}

In figures 6 and 7, we show that the loop spectra have different spectral properties from the Galactic diffuse spectra at 3.2--3.6 $\mu$m, which are characterized by the faint PAH 3.3 $\mu$m emission and the broad excess above the linear baseline. 
Figure 9 shows the relation between the ratio of the PAH 3.3 $\mu$m to the 9 $\mu$m band intensity and that of the 9 $\mu$m to the IRAS 100 $\mu$m band intensity for each observational point. In order to estimate robust measures of the PAH 3.3 $\mu$m intensity and its error from each spectrum in figure 7, we took the sum of the continuum-subtracted flux densities and the square root of the sum of the squares of their errors over the wavelength range of 3.20--3.29 $\mu$m, i.e., the shorter-wavelength side of the peak that is not affected by the sub-features, and doubled them. As for the 9 $\mu$m band intensities, except for data point (e) in figure 9, we multiply the surface brightness averaged over the area of the sub-slit aperture by the 9 $\mu$m band width of 18.9 THz \citep{Ona07} using the 9 $\mu$m emission map in figure 1. For data point (e), we calculate the 9 $\mu$m band intensity by combining the spectrum with the filter response curve in figure 2. To plot data point (e) and the others in the same graph, we apply the preliminary conversion factor from ADU to MJy sr$^{-1}$ for the 9 $\mu$m map and tentatively adopt a 10 \% level of the brightness as an error. For the IRAS 100 $\mu$m band, we similarly multiply the surface brightness at each position by the band width of 1.1 THz \citep{Neu84}. 

For the PAH 3.3 $\mu$m and the 9 $\mu$m intensity, we have to consider the reddening effect due to interstellar extinction. The interstellar extinction curves in the near- to mid-IR measured by ISO for the Galactic center region suggest $A_{\rm 3.3 \mu m}\sim 0.07\times A_{\rm V}$ mag and $A_{\rm 7.5 \mu m} \sim 0.04\times A_{\rm V}$ mag \citep{Lut96}. From the column density of neutral hydrogen atom gas, $N_{\rm H}$, to each direction \citep{Sta92} as well as the canonical relation $N_{\rm H}/A_{\rm V} = 2\times10^{21}$  cm$^{-2}$mag$^{-1}$ \citep{Boh78}, we estimate a dereddening factor, which is 1.3--1.9 for the 3.3 $\mu$m intensity and 1.2--1.5 for the 9.0 $\mu$m intensity, depending on the observational position. In figure 9, we plot results both corrected and uncorrected for the interstellar extinction. The figure clearly reveals the systematically lower ratios of the PAH 3.3 $\mu$m to the 9 $\mu$m band intensity for the loop spectra than for the Galactic diffuse spectra. The additional contribution of molecular gas to the estimate on $A_{\rm V}$ would make the difference even larger. In contrast, the ratios of the 9 $\mu$m to the 100 $\mu$m band intensity exhibit much smaller differences between these spectra, although they do show small decreases at the foot points of the loops as demonstrated in figures 4 and 5. Without the extinction corrections, the ratios of the 9 $\mu$m to the 100 $\mu$m intensity are 0.28--0.38 for the on-plane diffuse regions and 0.25--0.28 for the loop regions, which are consistent with typical values expected for the diffuse ISM by the standard dust model \citep{Li01}. It is therefore unlikely that unresolved low-luminosity old stars contribute much to the 9 $\mu$m emission in the loop regions \citep{Cha98}. Thus figure 9 shows the unusual faintness of the PAH 3.3 $\mu$m emission for the loop spectra, relative to the other PAH features dominating the 9 $\mu$m band.

In comparison to PAH emission features at longer wavelengths, the relative strength of the 3.3 $\mu$m emission is sensitive to PAHs of small sizes \citep{Sch93}. 
Generally, as compared with larger PAHs, small PAHs have much lower absorption efficiency for visible photons than for UV photons \citep{Sch93,Li02}.
Therefore, systematic difference in radiation hardness between the on-plane and off-plane regions could be responsible for the faintness in the PAH 3.3 $\mu$m emission of the loop spectra. The radiation hardness can be probed, for example, by the ratio of the [C{\small II}] 158 $\mu$m line to far-IR continuum emission. \citet{Nak98} derived the [C{\small II}]/far-IR ratio map of the Galactic center region, which indicated no systematic differences in radiation hardness between the on-plane and off-plane regions, although the Galactic center itself showed a global depression in [C{\small II}]/far-IR ratio along the Galactic plane.
Therefore it is rather unlikely that the unusual faintness of the PAH 3.3 $\mu$m emission for the loop spectra is caused by softer radiation fields in the off-plane regions. Another possibility is that the PAHs are locally shielded by very dense gas, which is also unlikely from the above discussion based on figure 8. We thus conclude that a significant fraction of small PAHs have been destroyed in the regions associated with loop 2. 

As for the 3.2--3.6 $\mu$m excess, if the broad feature in the spectrum of figure 6c is real, it can be emitted by aliphatic C-H stretching \citep{Dul81}. Small hydrocarbon particles may have been produced near the foot point of the molecular loop by shattering of larger carbonaceous grains \citep{Jon96}, while pre-existing small PAHs have been destroyed there. These particles may have not yet been graphitized by UV heating like in planetary nebulae as formation sites of PAHs. We have to await future follow-up observations to ensure the significance of the broad excess, because this is the only example we obtain from the spectrum with large systematic uncertainties.

\section{Conclusions}
We have created the AKARI 9 $\mu$m broad band image of an area of about $4^{\circ}\times3^{\circ}$ near the Galactic center including the CO molecular loops found with the NANTEN telescope (loops 1 and 2). We find the presence of 9 $\mu$m emission extended toward high Galactic latitudes, which spatially corresponds to the molecular loop structure. Near the foot points of the loops, the 9 $\mu$m emission is significantly suppressed as compared with the IRAS 100 $\mu$m emission, suggesting the efficient destruction of PAHs relative to submicron dust grains in these regions. We have derived near-IR spectra from the regions associated with loop 2 to compare their hydrocarbon spectral features with those typical of Galactic diffuse emission. We find that the loop spectra have different spectral properties from the Galactic diffuse spectra at 3.2--3.6 $\mu$m, which are characterized by the faint 3.3 $\mu$m aromatic emission and the broad excess above the linear baseline. The former represents the absence of very small PAHs, while the latter possibly suggests the dominance of aliphatic hydrocarbons. These suggest the destruction of very small PAHs and the shattering of carbonaceous grains. The CO radio observations showed that the violent motion and the shock heating of gas took place in the loops. Our IR photometric and spectroscopic results provide another piece of observational evidence for such high energetic phenomena in the loops.

\bigskip

We would express many thanks to an anonymous referee for giving us many valuable comments.
This work is based on observations made with AKARI. AKARI is a JAXA project with the participation of ESA. We thank the AKARI project manager, Hiroshi Murakami; part of the observations for the present study was carried out by using the Director's Time. This research is supported by a Grant-in-Aid for Scientific Research No. 22340043 from the Japan Society for the Promotion of Science, and the Nagoya University Global COE Program, ``Quest for Fundamental Principles in the Universe: from Particles to the Solar System and the Cosmos'', from the Ministry of Education, Culture, Sports, Science and Technology of Japan.


\end{document}